\documentclass[11pt]{article}

\usepackage{amsmath, amsfonts, amssymb, bm}
\usepackage{graphicx,psfrag,epsf, float}
\usepackage{enumerate,titlesec, color, rotating}
\RequirePackage[authoryear]{natbib}
\usepackage{lineno}
\usepackage{url} 
\usepackage{soul}
\usepackage{subcaption}
\usepackage{booktabs}
\usepackage{tikz, tkz-graph, xcolor}
\usepackage[ruled,linesnumbered]{algorithm2e}

\newcommand{\mH}{{\mathcal H}}

\newcommand{\T}{{\mathcal T}}

\newcommand{\hS}{{\widehat S}}
\newcommand{\sumi}{{\sum_{i=1}^n}}

\newcommand{\sumk}{{\sum_{k=1}^n}}
\newcommand{\sumb}{{\sum_{b=1}^B}}

\newcommand{\bX}{{\bm X}}
\newcommand{\bW}{{\bm W}}

\newcommand{\bx}{{\bm x}}
\newcommand{\bz}{{\bm z}}
\newcommand{\bZ}{{\bm Z}}

\newcommand{\CON}{{\rm CON}}
\newcommand{\TPR}{{\rm TPR}}
\newcommand{\FPR}{{\rm FPR}}
\newcommand{\NA}{{\rm NA}}

\newcommand{\DeltaL}{{\Delta_{\rm L}}}
 
\newcommand{\bl}[1]{\textcolor{blue}{#1}}

\definecolor{firstposColor}{HTML}{FFb2b2}
\definecolor{chronicColor}{HTML}{FF3232}
\definecolor{mucoidColor}{HTML}{990000}
\definecolor{MDRColor}{HTML}{934c93}

\newcommand\crule[3][black]{\textcolor{#1}{\rule{#2}{#3}}}

\newcommand{\lineCir}[2][]{\tikz[baseline=-.5ex]{\draw[#1, line width = 2pt](-.4,0) --(.4,0);\draw[#1, fill = #1] (0, 0) circle (#2);}}
\newcommand{\dashlineCir}[2][]{\tikz[baseline=-.5ex]{\draw[#1, dotted, line width = 2pt](-.4,0) --(.4,0);\draw[#1, fill = #1] (0, 0) circle (#2);}}

\newcommand{\TL}{{T_{\rm L}}}
\newcommand{\YL}{{Y_{\rm L}}}
\newcommand{\YLi}{{Y_{{\rm L}i}}}

\DeclareMathOperator{\logit}{logit\!}

\newcommand\blfootnote[1]{%
  \begingroup
  \renewcommand\thefootnote{}\footnote{#1}%
  \addtocounter{footnote}{-1}%
  \endgroup
}

\begin{document}

  \title{Dynamic Risk Prediction Triggered by Intermediate Events Using Survival Tree Ensembles}
	\date{}
	\author{Yifei Sun, Sy Han Chiou, Colin O.Wu, Meghan McGarry, and Chiung-Yu Huang\blfootnote{Yifei Sun is Assistant Professor, Department of Biostatistics, Columbia University Mailman School of Public Health, New York, NY 10032 (Email: \emph{ys3072@cumc.columbia.edu}). 
	Sy Han Chiou is Assistant Professor, Department of Mathematical Sciences, University of Texas at Dallas (Email: \emph{schiou@utdallas.edu}).
	Meghan McGarry is Assistant Professor, Department of Pediatrics, School of Medicine, University of California San Francisco (Email: \emph{Meghan.McGarry@ucsf.edu}).
	Colin O. Wu is Mathematical Statistician, National Heart, Lung, and Blood Institute, National Institutes of Health (Email: \emph{wuc@nhlbi.nih.gov}). 
	Chiung-Yu Huang is Professor, Department of Epidemiology and Biostatistics, School of Medicine, University of California San Francisco, San Francisco, CA 94158 (Email: \emph{ChiungYu.Huang@ucsf.edu}).}}
	\maketitle

  \begin{abstract} 
    With the availability of massive amounts of data from electronic health records and registry databases, incorporating time-varying patient information to improve risk prediction has attracted great attention. To exploit the growing amount of predictor information over time, we develop a unified framework for landmark prediction using survival tree ensembles, where an updated prediction can be performed when new information becomes available. Compared to conventional landmark prediction with fixed landmark times, our methods allow the landmark times to be subject-specific and triggered by an intermediate clinical event. Moreover, the nonparametric approach circumvents the thorny issue of model incompatibility at different landmark times. In our framework, both the longitudinal predictors and the event time outcome are subject to right censoring, and thus existing tree-based approaches cannot be directly applied. To tackle the analytical challenges, we propose a risk-set-based ensemble procedure by averaging martingale estimating equations from individual trees. Extensive simulation studies are conducted to evaluate the performance of our methods. The methods are applied to the Cystic Fibrosis Patient Registry (CFFPR) data to perform dynamic prediction of lung disease in cystic fibrosis patients and to identify important prognosis factors.
  \end{abstract}

KEY WORDS: Dynamic prediction, landmark analysis, multi-state model, survival tree, time-dependent predictors. 

\def\spacingset#1{\renewcommand{\baselinestretch}%
	{#1}\small\normalsize} \spacingset{1}
\spacingset{1.45}

\section{Introduction}

Cystic fibrosis (CF) is a genetic disease characterized by a progressive, irreversible decline in lung function caused by chronic microbial infections of the airways. Despite recent advances in diagnosis and treatment, the burden of CF care remains high, and most patients succumb to respiratory failure. There is currently no cure for CF, so early prevention of lung disease for high-risk patients are essential for successful disease management. The goal of this research is to develop flexible and accurate event risk prediction algorithms for abnormal lung function in pediatric CF patients by exploiting the {rich longitudinal} information made available by the Cystic Fibrosis Foundation Patient Registry (CFFPR). 

The CFFPR is a large electronic health record database that collects encounter-based records of over 300 unique variables on patients from over 120 accredited CF care centers in the United States \citep{knapp2016cystic}. The CFFPR contains detailed information on potential risk factors, including symptoms, pulmonary infections, medications, test results, and medical history. Analyses of CFFPR suggested that the variability in spirometry measurements over time is highly predictive of subsequent lung function decline \citep{morgan2016forced}. Moreover, the acquisition of chronic, mucoid, or multidrug-resistant subtypes of \emph{Pseudomonas aeruginosa} (PA) leads to more severe pulmonary disease, accelerating the decline in lung function \citep{meghan2020early}. In this paper, the event of interest is the progressive loss of lung function, defined as the first time that the percent predicted forced expiratory volume in 1 second (ppFEV1) drops below 80\% in CFFPR. Since risk factors such as weight and height in pediatric patients can change substantially over time, models with baseline predictors have limited potential for long-term prognosis. Incorporating repeated measurements and intermediate clinical events would reflect ongoing CF management and result in more accurate prediction.

To incorporate the longitudinal patient information in risk prediction, one major approach is joint modeling \citep[see, for example,][]{rizopoulos2011dynamic,taylor2013real}. Under the joint modeling framework, a longitudinal submodel for the time-dependent variables and a survival submodel for the time-to-event outcome are postulated; the sub-models are typically linked via latent variables. Such a model formulation provides a complete specification of the joint distribution, based on which the survival probability given the history of longitudinal measurements can be derived. Most joint modeling methods consider a single, continuous time-dependent variable.
Although attempts have been made to incorporate multiple time-dependent predictor variables \citep{proust2016joint,wang2017dynamic}, correct specification of the model forms for all the time-dependent covariates and their associations with the event outcome remains a major challenge. Moreover, it is not clear how existing joint modeling approaches can further incorporate the information on the multiple intermediate events, such as the acquisition of different subtypes of PA, in risk prediction.

Another major approach that can account for longitudinal predictors is landmark analysis, where models are constructed at pre-specified landmark times to predict the event risk in a future time interval. For example, at each landmark time, one may postulate a working Cox model with appropriate summaries of the covariate history up to the landmark time (e.g., last observed values) as predictors and then fit the Cox model using data from subjects who are at risk of the event. The estimation can either be performed using a separate model at each landmark time point or a supermodel for all landmark time points \citep{van2007dynamic,van2008dynamic,van2011dynamic}. This way, multiple and mixed type time-dependent predictors can be easily incorporated. Moreover, to better exploit the repeated measurements and to handle measurement errors, one may also consider a two-stage landmark approach \citep{rizopoulos2017dynamic,sweeting2017use,ferrer2019individual}: in the first step, mixed-effects models are used to model the longitudinal predictors; in the second step, functions of the best linear unbiased prediction (BLUP) estimator of the random effects are included as predictors of the landmark Cox model.
Other than Cox models, \cite{parast2012landmark} considered time-varying coefficient models to incorporate a single intermediate event and multiple biomarker measurements. \cite{zheng2005partly}, \cite{maziarz2017longitudinal}, and \cite{zhu2019landmark} further considered the impact of informative observation times of repeated measurements on future risk.

Direct application of the existing landmark analysis method to the CFFPR data may not be ideal for the following reasons: first, imposing semiparametric working models at different landmark times may result in incompatible models and inconsistent predictions \citep{jewell1993framework, rizopoulos2017dynamic}. In other words,  a joint distribution of predictors and event times that satisfies the models at all the landmark times simultaneously may not exist. Second, the specification of how the predictor history affects the future event risk may require deep clinical insight. For example, researchers have shown that various summaries of the repeated measurements, including the variability \citep{morgan2016forced}, the rate of change \citep{mannino2006lung}, and the area under the trajectory curve \citep{domanski2020time}, can serve as important predictors of disease risks, while the last observed value has been commonly used in the statistical literature. Therefore, nonparametric statistical learning methods are appealing in landmark prediction, because they require minimal model assumptions and have the potential to deal with a large number of complicated predictors. \cite{tanner2021dynamic} applied super learners for landmark prediction in CF patients, where discrete time survival analysis were conducted via the use of machine learning algorithms for binary outcomes.

In this paper, we propose a unified framework for landmark prediction using survival tree ensembles, where the landmark times can be subject-specific. A subject-specific landmark can be defined by an intermediate clinical event that modifies patients' risk profiles and triggers the need for an updated evaluation of future risk. In our application, the acquisition of chronic PA usually leads to accelerated deterioration in the pulmonary function and serves as a natural landmark. When the landmark time is random, the number of observed predictors at the landmark time often varies across subjects, creating analytical challenges in fully utilizing the available information. Moreover, unlike static risk prediction models where baseline predictors are completely observed, the observation of the time-dependent predictors is further subject to right censoring. To tackle these problems, we propose a risk-set-based approach to handle the possibly censored predictors. To avoid the instability issue of a single tree, we propose a novel ensemble procedure based on averaging unbiased martingale estimating equations derived from individual trees. Our ensemble method is different from existing ensemble methods that directly average the cumulative hazard predictions and has strong empirical performances in dealing with censored data.

The rest of this article is organized as follows. In Section 2, we introduce a landmark prediction framework that incorporates the repeated measurements and intermediate events. In Section 3, we propose tree-based ensemble methods to deal with censored predictors and outcomes. We propose a concordance measure to evaluate the prediction performance in Section 4 and define a permutation variable importance measure in Section 5. The proposed methods are evaluated by extensive simulation studies in Section 6 and are applied to the CFFPR data in Section 7. We conclude the paper with a discussion in Section 8.

\section{Model Setup}
\label{sect:model}

In contrast to static risk prediction methods that output a conditional survival function given baseline predictors, dynamic landmark prediction focuses on the survival function conditioning on the predictor history up to the landmark time. Since history information involves complicated stochastic processes, challenges arise as to how to partition the history processes when applying tree-based methods. In what follows, we first introduce a generalized definition of the landmark survival function, starting from either a fixed or subject-specific landmark time. We then express the history information as a fixed-length predictor vector on which recursive partition can be applied.

Denote by $T$ a continuous failure event time and by $\TL$ a landmark time. The landmark is selected a priori and is usually clinically meaningful. We allow $\TL$ to be either fixed or subject-specific. We focus on the subpopulation that is free of the failure event at $\TL$ and predict the risk after $\TL$. Denote by $\bZ$ the baseline predictors and denote by $\mH(t)$ other information observed on $[0,t]$. Our goal is to predict the probability conditioning on all the available information up to $\TL$, that is,
\begin{align}
\label{pr}
P(T-\TL \ge t\mid  T\ge \TL,  \TL , \mH(\TL) , \bZ).
\end{align}
To illustrate the observed history $\mH(t)$, we consider two types of predictors that are available in the CFFPR data. 
The first type of predictors is repeated measurements of time-dependent variables such as weight and ppFEV1. It is worthwhile to point out that both internal and external time-dependent predictors can be included, as only their history up to $\TL$ will be used.
We denote this type of predictors by $\bW(t)$, a $q$-dimensional vector of time-dependent variables, and assume $\bW(\cdot)$ is available at fixed time points $t_1,t_2,\ldots,t_K$. The observed history up to $t$ is
$\mH_W(t) = \{ \bW(s)d O(s), 0< s\le t\}$, where $O(t)$ is a counting process that jumps by one when $\bW(\cdot)$ is measured (i.e., $dO(t_k) = 1$ for $k = 1,\ldots,K$). 
The second type of predictors is the timings of intermediate clinical events such as pseudomonas infections. Denote by $U_j$ the time to the $j$th intermediate event, $j = 1,\ldots, J$. The observed history up to $t$ is 
$\mH_U(t) = \{ I(U_j\le s), 0< s\le t, j = 1,\ldots,J\}$. Collectively, we have a system of history processes $\mH(t) = (\mH_W(t), \mH_U(t))$.

\begin{figure}
  \centering
  \begin{subfigure}[t]{0.45\textwidth}
    \resizebox{\textwidth}{!}{
      \begin{tikzpicture}
        \draw[line width = .6mm, ->] (4, 0) -- (11.7, 0);
        \draw[line width = .75mm, blue!80!gray] (7, -.15) -- (7, .15);
        \node[blue!80!gray] at (7, -.4) {\normalsize $\boldsymbol{t_1 = 7}$};
        \draw[line width = .75mm, black!80] (4.84, -.15) -- (4.84, .15);
        \node[black] at (4.84, -.4) {\normalsize cPA};
        \draw[line width = 3.5mm, cyan!60!gray, opacity = 0.45] (7, 0) -- (11.7, 0);
        \node[black!60!green] at (7.7, .4) {\normalsize $\boldsymbol{W(t_1) = 19.7}$ kg};
        \node[black!60!green] at (4.84, .4) {\normalsize $\boldsymbol{U_1 = 4.8}$};        
      \end{tikzpicture}}
    \caption{Subject 1, landmark at $t_1 = 7$.}
    \label{fig:ex1b}
  \end{subfigure}
    \begin{subfigure}[t]{0.45\textwidth}
    \resizebox{\textwidth}{!}{
      \begin{tikzpicture}
        \draw[line width = .6mm, ->] (4,0) -- (11.7, 0);
        \draw[line width = .75mm, blue!80!gray] (7, -.15) -- (7, .15);
        \node[blue!80!gray] at (7, -.4) {\normalsize $\boldsymbol{t_1 = 7}$};
        \draw[line width = .75mm, black!50] (10.22, -.15) -- (10.22, .15);
        \node[black!20] at (10.22, -.4) {\normalsize cPA};
        \draw[line width = 3.5mm, cyan!60!gray, opacity = 0.45] (7, 0) -- (11.7, 0);
        \node[black!60!green] at (7.7, .4) {\normalsize $\boldsymbol{W(t_1) = 21.5}$ kg};
        \node[black!60!green] at (10.22, .4) {\normalsize $\boldsymbol{U_1 > t_1}$};
      \end{tikzpicture}}
    \caption{Subject 2, landmark at $t_1 = 7$.}
    \label{fig:ex1a}
  \end{subfigure}
  \begin{subfigure}[t]{0.45\textwidth}
    \resizebox{\textwidth}{!}{
      \begin{tikzpicture}
        \draw[line width = .6mm, ->] (4, 0) -- (11.7, 0);
        \draw[line width = .75mm, black!80] (7, -.15) -- (7, .15);
        \node[black!80] at (7, -.4) {\normalsize $\boldsymbol{t_1 = 7}$};
        \draw[line width = .75mm, blue!80!gray] (4.84, -.15) -- (4.84, .15);
        \node[blue!80!gray] at (4.84, -.4) {\normalsize cPA};
        \draw[line width = 3.5mm, cyan!60!gray, opacity = 0.45] (4.84, 0) -- (11.7, 0);
        \node[black!20] at (7.7, .4) {\normalsize $\boldsymbol{W(t_1) = 19.7}$ kg};
        \node[black!60!green] at (4.84, .4) {\normalsize $\boldsymbol{U_1 = 4.8}$};
      \end{tikzpicture}}
    \caption{Subject 1, landmark at $U_1$.}
    \label{fig:ex1d}
  \end{subfigure}  
  \begin{subfigure}[t]{0.45\textwidth}
    \resizebox{\textwidth}{!}{
      \begin{tikzpicture}
        \draw[line width = .6mm, ->] (4, 0) -- (11.7, 0);
        \draw[line width = .75mm, black!80] (7, -.15) -- (7, .15);
        \node[black!80] at (7, -.4) {\normalsize $\boldsymbol{t_1 = 7}$};
        \draw[line width = .75mm, blue!80!gray] (10.22, -.15) -- (10.22, .15);
        \node[blue!80!gray] at (10.22, -.4) {\normalsize cPA};
        \draw[line width = 3.5mm, cyan!60!gray, opacity = 0.45] (10.22, 0) -- (11.7, 0);
        \node[black!60!green] at (7.7, .4) {\normalsize $\boldsymbol{W(t_1) = 21.5}$ kg};
        \node[black!60!green] at (10.55, .4) {\normalsize $\boldsymbol{U_1 = 10.2}$};
      \end{tikzpicture}}    
    \caption{Subject 2, landmark at $U_1$.}
    \label{fig:ex1c}
  \end{subfigure}
  \caption{Illustration of fixed and random landmark times.
    \crule[blue!80!gray]{.05cm}{.25cm} marks the landmark time point;
    \crule[black!60!green]{.25cm}{.25cm} marks the available information at the landmark time;
    \crule[black!20]{.25cm}{.25cm} marks the unavailable information at the landmark time;
    \crule[cyan!60!gray]{.25cm}{.25cm} marks the target prediction interval.
  }
  \label{fig:ex1}
\end{figure}
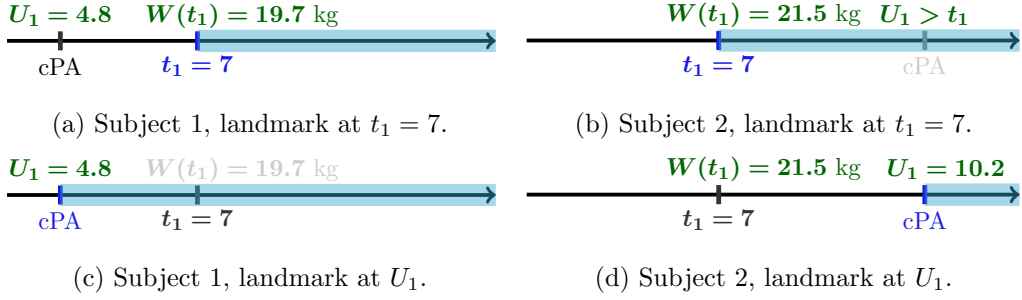

In our framework, both $t_k$ and $U_j$ can serve as landmark times.
Due to the stochastic nature of $U_j$, the order of $\{t_k, U_j, k = 1,\ldots, K, j = 1,\ldots, J\}$ can not be pre-determined. As a result, the number of available predictors at a given landmark time can vary across subjects. 
For illustration, we consider one fixed time point $t_1$ at age 7 and one intermediate event chronic PA (cPA), of which the occurrence time is denoted by $U_1$. Figure~\ref{fig:ex1} depicts the observed data of two study subjects. At $t_1 = 7$, subject 1 has experienced cPA ($U_1 = 4.8 \le t_1$), while 
subject 2 remains free of cPA ($U_1 = 10.2 > t_1$). 
Let $\bW(t)$ be the body weight measured at time $t$ (i.e., $q=1$). The probabilities of interest are given as follows:

\begin{enumerate}
\item[(I)] At a fixed landmark time $\TL = t_1$, we predict the risk at $t_1+t$, $t>0$, among those who are at risk, that is, $T\ge t_1$. Note that subjects in the risk set may or may not have experienced the intermediate event prior to $t_1$. Given $\bW(t_1)$ and the partially observed $U_1$, the conditional survival probability \eqref{pr} can be reexpressed as
\begin{align*}
\begin{cases}
    P(T\ge t + t_{1} \mid T\ge t_{1} , \bZ , \bW(t_1) , U_1 ), & \text{if~} U_1 \le t_1, \\
    P(T\ge t + t_{1}\mid T\ge t_{1} , \bZ, \bW(t_1) , U_1 > t_{1}), & \text{otherwise}.
\end{cases}
\end{align*}
In other words, at $t_1$, we output the former for subjects who experience the intermediate event prior to $t_1$ (subject 1, Figure~\ref{fig:ex1b}), while output the latter for others (subject 2, Figure~\ref{fig:ex1a}).
\item[(II)] At a random landmark time $\TL = U_1$, we predict the risk for subjects who have experienced the intermediate event and are free of the failure event. The predictor value $\bW(t_1)$ is available only if $U_1 \ge t_1$. In this case, we predict 
  \begin{align*}
\begin{cases}
      P(T\ge t + U_1\mid T\ge U_1, \bZ , U_1 ), & \text{if~} U_1 \le t_1, \\
      P(T\ge t + U_1\mid T\ge U_1, \bZ , \bW(t_1) , U_1 ),  & \text{otherwise}.
\end{cases}
  \end{align*}
Therefore, at $U_1$, we output the former for subjects whose $\bW(t_1)$ is observed after $U_1$ (Subject 1, Figure~\ref{fig:ex1d}), while output the latter for others (Subject 2, Figure~\ref{fig:ex1c}). 
\end{enumerate}

In the above example, the observed predictors vary across subjects. We then represent the history $\mH(\TL)$ as a vector with a fixed length, so that tree-based methods can be applied to estimate the probability in \eqref{pr}. Define the complete predictors $\bX = \{\bW(t_1), \ldots, \bW(t_K), U_1,\ldots,U_J\}$. The information in $\bX$ may not be fully available at a given landmark time. We define the available information up to $t$ by $\bX(t) = \{\bW(t_1,t), \ldots,$ $\bW(t_K,t), U_1(t),\ldots,U_J(t)\}$, where 
\begin{align*}
    \bW(t_k,t) = \begin{cases}
    \bW(t_k), & \text{~if~} t_k \le t,\\
    {\textbf{NA}_q}, & \text{~otherwise},
    \end{cases}\hspace{0.3in}
    U_j(t) = \begin{cases}
    U_j, & \text{~if~} U_j \le t,\\
    t^+, & \text{~otherwise}.
    \end{cases}
\end{align*}
By a slight abuse of notation, we write $U_j(t) = t^+$ if $U_j > t$ and  $\bW(t_k,t) = \textbf{NA}_q$ if $t_k > t$, with $\textbf{NA}_q \overset{\rm def}{=} (\rm NA, \ldots, NA)$ denoting a non-numeric $q$-dimensional vector. Here an NA value indicates that the covariate value is collected after the landmark time and thus should not be used for constructing prediction models. In other words, under our setting NA is treated as an attribute rather than missing data, as the target probability is not conditional on $\bW(t_k)$ for $t_k>\TL$. The covariate history $\mH(\TL)$ can then be expressed as $\bX(\TL)$, which is a $(qK+J)$-dimensional vector. This way,  a covariate not being observed is predictive of the outcome, and the target survival  probability function can be expressed as follows:
\begin{align}
\label{pr3}
S(t\mid a, \bz, \bx) = P(T\ge t+a \mid T\ge \TL = a, \bZ = \bz, \bX(\TL) = \bx) .
\end{align}
In the example depicted in Figure~\ref{fig:ex1}, we have $\bX(t) = \{\bW(t_1,t),U_1(t) \}$. For $\TL = t_1$, the predictor values in Figure~\ref{fig:ex1b} and \ref{fig:ex1a} correspond to $\bx = (19.7, 4.8)$ and $\bx = (21.5, 7^+)$, respectively.
For $\TL = U_1$, the predictor values in Figures~\ref{fig:ex1d} and \ref{fig:ex1c} correspond to $\bx = ({\rm NA}, 4.8)$ and $\bx = (21.5, 10.2)$, respectively. Since the information at $\TL$ involves left-bounded intervals and non-numeric values, applying semiparametric methods to estimate $S(t\mid a,\bz,\bx)$ is challenging. 
Hence we propose tree-based methods to handle partially observed predictors.

\section{Survival trees and ensembles for landmark prediction}
At a given landmark time, we build a tree-based model to predict future event risk. Survival trees are popular nonparametric tools for risk prediction. The original survival trees take baseline predictors as input variables and output the survival probability conditioning on the baseline predictors \citep[see, for example,][] {gordon1985tree,ciampi1986stratification,segal1988regression,davis1989exponential,leblanc1992relative,leblanc1993survival,zhang1995splitting,molinaro2004tree,steingrimsson2016doubly}, and ensemble methods have been applied to address the instability issue of a single tree \citep{hothorn2004bagging,hothorn2006survival,ishwaran2008random,zhu2012recursively,steingrimsson2018censoring}. However, existing methods may not be directly applied, because the predictors in $\bX$ are not completely observed at $\TL$, and the available predictors $(\TL,\bX(\TL))$ are subject to right censoring. In the absence of censoring, we introduce a partition scheme for subjects who are event-free at the landmark time in Section \ref{sect:part}. To handle censored data, we propose risk-set methods to estimate the partition-based landmark survival probability in Section \ref{sect:censor} and propose an ensemble procedure in Section \ref{sect:ensemble}.

\subsection{Partition on partially observed predictors at the landmark time}
\label{sect:part}

A tree partitions the predictor space into disjoint subsets termed terminal nodes and assigns the same survival prediction for subjects that enter the same terminal node. In dynamic risk prediction, the population of interest is subjects who remain event-free at the landmark time, that is, those with $T\ge\TL$. We use $\T = \{\tau_1,\tau_2,\ldots,\tau_M \}$ to denote a partition on the sample space of $(\TL,\bZ,\bX(\TL))\mid T\ge \TL$, where $\tau_m, m = 1,\ldots,M$, are the terminal nodes. The terminal nodes are formed recursively using binary partitions by asking a sequence of yes-or-no questions.
Existing implementations of trees usually do not handle mixtures of numeric and nominal variables.
Since a variable in $\bX(\TL)$ may take either numeric/ordinal values or NA, the conventional partition scheme needs to be extended. When a variable in $\bW(t_k)$ is nominal, its counterpart in $\bW(t_k,\TL)$ is also nominal and can be split applying existing approaches. In what follows, we focus on the case where the longitudinal marker measurements $\bW(t_k)$ are numeric/ordinal.

When $\TL$ is random, the partition is based on the variables in $(\TL,\bZ,\bX(\TL))$. We consider the following situations:
\begin{enumerate}
    \item[(R1)] When $\bW(t_k,\TL)$'s are {the} splitting variables, conventional splitting rules may not be directly applied because they take NA values when $t_k>\TL$.
    Supposed $W$ is an element of $\bW(t_k,\TL)$ and $c$ is a cutoff value. We consider two possible splits,
    (a) \{$W > c$\} versus \{$W \le c$ or $W = \NA$\}  and (b)  \{$W > c$ or $W = \NA$\} versus \{$W \le c$\},
    and select the split that yields a larger improvement in the splitting criterion. We note that $W = \rm NA$ (i.e., $\TL < t_k$) is treated as an attribute rather than missing data.
    \item[(R2)] 
    Conventional splitting rules cannot be directly applied to $U_j(\TL)$, because $U_j(\TL)$ can be $\TL^+$ and the support of $U_j(\TL)$ is not an ordered set. To tackle this problem, we employ a set of transformed predictors $\{U_1(\TL)/\TL,\ldots,U_J(\TL)/\TL\}$, which, together with $\TL$, contains the same information as $\{U_1(\TL),\ldots,U_J(\TL),\TL\}$.
    Then the rule $U_j(\TL)/\TL > c$ can be applied for $c\in(0,1]$, and $U_j(\TL)/\TL > 1$ is equivalent to $U_j(\TL)=\TL^+$. 
\end{enumerate}

The splitting rules in (R1) and (R2) can be implemented by transforming $\bX(\TL)$ to a numeric/ordinal vector and passing the transformed predictors into a tree algorithm using conventional partition rules: First, we create two features for each element $W$ in $\bW(t_k,\TL)$. The two features, denoted by $W^+$ and $W^-$, take the same value as $W$ when $W\neq {\NA}$ (i.e., $t_k>\TL$), and take extreme values $M$ and $-M$ otherwise, where $M$ is a large positive number outside the possible range of predictor values. Instead of using $W$, we use $W^+$ and $W^-$ as candidate variables for splitting. The partition ``\{$W^+ > c$\} versus \{$W^+ \le c$\}'' is equivalent to the type (b) partition ``\{$W>c$ or $W = {\rm NA}$\} versus \{$W\le c$\}'', because observations {satisfying} $W={\rm NA}$ or $W>c$ are assigned to the same child node; similarly, splitting based on $W^-$ yields type (a) partitions. As an example, if landmark is cPA and $W$ denotes the observed age-7 weight at the landmark time, then the rule ``Is $W^+>20$?'' is equivalent to ``Is weight at age 7 greater than 20kg or does the subject have cPA before age 7?''. Implementing such a partition rule is equivalent to the ``missings together'' approach \citep{zhang1996tree} and missingness incorporated in attributes \citep{twala2008good}.
Second, for each transformed intermediate event time $U_j(\TL)/\TL (j = 1,\ldots,J)$, we replace the value $1^+$ with $M$. The details are described in Algorithm 1 (see Table \ref{tab:a1}). The partition scheme also apply to the trivial case of fixed landmark times. Figure \ref{fig:ex2} illustrates the proposed data preprocessing procedure.

\begin{figure}[th]
\setlength{\tabcolsep}{2.0pt}
\renewcommand{\arraystretch}{1.5}
  \begin{tabular}{cclcccccc}
    \toprule\\[-5ex]
    Landmark &&& At-risk status & \multicolumn{2}{c}{$\bW(t_1, \TL)$} & \multicolumn{2}{c}{$U_1(\TL)$}\\[-1ex]
    \cmidrule(l){5-6}\cmidrule(l){7-8}\\[-5ex]
    time $\TL$ & ID & Observed data & $I(T\ge \TL)$ & Observed & Processed & Observed & Processed \\
    \midrule
    $t_1 = 7$ & 1 & 
            \resizebox{.33\textwidth}{!}{
            \begin{tikzpicture}[baseline = 0]
              \def\a {9.7}; 
              \def\b {0.15}; 
              \draw[line width = .6mm, -, white] (4, 0) -- (12.3, 0); 
              \draw[line width = .6mm, -, white] (4, -.5) -- (4, .5); 
              \draw[line width = .6mm, -] (4, 0) -- (\a, 0);
              \fill[black] (\a - \b, -\b) rectangle (\a + \b, \b);
              \draw[line width = .75mm, blue!80!gray] (7, -.15) -- (7, .15);
              \node[blue!80!gray] at (7, -.4) {\normalsize $\boldsymbol{t_1 = 7}$};
              \draw[line width = .75mm, black!80] (4.84, -.15) -- (4.84, .15);
              \node[black!60!green] at (7.7, .4) {\normalsize $\boldsymbol{W(t_1) = 19.7}$ kg};
              \node at (\a, -.4)
              {\normalsize $\boldsymbol{T = \a}$};
              \node[black!60!green] at (4.84, .4) {\normalsize $\boldsymbol{U_1 = 4.8}$};        
            \end{tikzpicture}} &
                                 1 & 19.7 & (19.7, 19.7) & 4.8 & 4.8 / 7 \\
          & 2 &
                \resizebox{.33\textwidth}{!}{
                \begin{tikzpicture}[baseline = 0]
                  \def\a {11.7}; 
                  \def\b {0.15}; 
                  \draw[line width = .6mm, -, white] (4, 0) -- (12.3, 0); 
                  \draw[line width = .6mm, -, white] (4, -.5) -- (4, .5); 
                  \draw[line width = .6mm, -] (4, 0) -- (\a - \b, 0);
                  \draw[line width = .6mm] (\a - \b, -\b) rectangle (\a + \b, \b);
                  \fill[black] (\a - \b, -\b) rectangle (\a + \b, \b);
                  \draw[line width = .75mm, blue!80!gray] (7, -.15) -- (7, .15);
                  \node[blue!80!gray] at (7, -.4) {\normalsize $\boldsymbol{t_1 = 7}$};
                  \draw[line width = .75mm, black!50] (10.22, -.15) -- (10.22, .15);
                  \node[black!60!green] at (7.7, .4) {\normalsize $\boldsymbol{W(t_1) = 21.5}$ kg};
                  \node[black!60!green] at (10.22, .4) {\normalsize $\boldsymbol{U_1 > t_1}$};
              \node at (\a, -.4)
              {\normalsize $\boldsymbol{T = \a}$};                \end{tikzpicture}} &
                                     1 & 21.5 & (21.5, 21.5) & $7^+$ & $M$ \\
          & 3 &
                \resizebox{.33\textwidth}{!}{
                \begin{tikzpicture}[baseline = 0]
                  \def\a {5.7}; 
                  \def\b {0.15}; 
                  \draw[line width = .6mm, -, white] (4, 0) -- (12.3, 0); 
                  \draw[line width = .6mm, -, white] (4, -.5) -- (4, .5); 
                  \draw[line width = .6mm, -] (4, 0) -- (\a - \b, 0);
                  \fill[black] (\a - \b, -\b) rectangle (\a + \b, \b);
              \node at (\a, -.4)
              {\normalsize $\boldsymbol{T = \a}$};
                \end{tikzpicture}} & 
                                     0 & - & -  & - & - \\
    [1ex]
    \midrule
    $U_1$ & 1 &
                \resizebox{.33\textwidth}{!}{
                \begin{tikzpicture}[baseline = 0]
                  \def\a {9.7}; 
                  \def\b {0.15}; 
                  \draw[line width = .6mm, -, white] (4, 0) -- (12.3, 0); 
                  \draw[line width = .6mm, -, white] (4, -.5) -- (4, .5); 
                  \draw[line width = .6mm, -] (4, 0) -- (\a, 0);
                  \fill[black] (\a - \b, -\b) rectangle (\a + \b, \b);
                  \draw[line width = .75mm, blue!80!gray] (4.84, -.15) -- (4.84, .15);
                  \node[blue!80!gray] at (4.84, -.4) {\normalsize cPA};
                  \node[black!60!green] at (4.84, .4) {\normalsize $\boldsymbol{U_1 = 4.8}$};
              \node at (\a, -.4)
              {\normalsize $\boldsymbol{T = \a}$};
                \end{tikzpicture}}
                           & 1 & \rm NA & ({$M$}, {$-M$}) & 4.8 & 4.8 / 4.8 \\
          & 2 & 
                \resizebox{.33\textwidth}{!}{
                \begin{tikzpicture}[baseline = 0]
                  \def\a {11.7}; 
                  \def\b {0.15}; 
                  \draw[line width = .6mm, -, white] (4, 0) -- (12.3, 0); 
                  \draw[line width = .6mm, -, white] (4, -.5) -- (4, .5); 
                  \draw[line width = .6mm, -] (4, 0) -- (\a - \b, 0);
                  \fill[black] (\a - \b, -\b) rectangle (\a + \b, \b);
                  \draw[line width = .75mm, blue!80!gray] (10.22, -.15) -- (10.22, .15);
                  \node[blue!80!gray] at (10.22, -.4) {\normalsize cPA};
                  \node[black!60!green] at (7.7, .4) {\normalsize $\boldsymbol{W(t_1) = 21.5}$ kg};
                  \node[black!60!green] at (10.55, .4) {\normalsize $\boldsymbol{U_1 = 10.2}$};
              \node at (\a, -.4)
              {\normalsize $\boldsymbol{T = \a}$};
                \end{tikzpicture}} &
                                     1 & 21.5 & (21.5, 21.5) & 10.2 & 10.2 / 10.2 \\
          & 3 &
                \resizebox{.33\textwidth}{!}{
                \begin{tikzpicture}[baseline = 0]
                  \def\a {5.7}; 
                  \def\b {0.15}; 
                  \draw[line width = .6mm, -, white] (4, 0) -- (12.3, 0); 
                  \draw[line width = .6mm, -, white] (4, -.5) -- (4, .5); 
                  \draw[line width = .6mm, -] (4, 0) -- (\a - \b, 0);
                  \fill[black] (\a - \b, -\b) rectangle (\a + \b, \b);
              \node at (\a, -.4)
              {\normalsize $\boldsymbol{T = \a}$};
                \end{tikzpicture}} &
                                     0 & - & -  & - & - \\
    \bottomrule
  \end{tabular}
  \begin{flushleft}
  \caption{Illustration of observed data and preprocessed data from three subjects at fixed and random landmark times. At each landmark time $\TL$, only subjects with $T\ge \TL$ (subjects 1 and 2) are of interest. For each predictor that can take the value NA (e.g., $\bW(t_1,\TL)$), we create two features that take extreme values $M$ and $-M$ if the predictor is not observed. For partially observed intermediate events (e.g., $U_1(\TL)$), we first divide the event time by $\TL$ and replace the value $1^+$ with $M$. After preprocessing, zero-variance features and duplicate features can be removed before running the tree algorithm.}
  \label{fig:ex2}
  \end{flushleft}
\end{figure}

The partition scheme described above guarantees that each individual has a well-defined pathway to determine its node membership. Given such a partition $\T$, one can define a partition function $l_{\T}(a,\bz,\bx)$, which returns the terminal node in $\T$ that contains $(a,\bz,\bx)$. We define the following partition-based survival function,
\begin{align}
\label{st}
    S_{\T}(t\mid a,\bz,\bx) = P( T\ge t+a\mid T\ge \TL, (\TL,\bZ,\bX(\TL)) \in l_{\T}(a,\bz,\bx)   ).
\end{align}
The probability $S_{\T}(t\mid a,\bz,\bx)$ approximates the target function $S(t\mid a,\bz,\bx)$.

\subsection{Partition-based estimation at the landmark time}
\label{sect:censor}
The follow-up of a subject can be terminated due to loss to follow-up or study end. 
We now consider estimating $S_\T(t\mid a,\bz,\bx)$ with censored data. Denote by $C$ the censoring time and assume $C$ is independent of $(\TL,T,\bX)$ given $\bZ$. We follow the convention to define $Y = \min(T,C)$ and $\Delta = I(T \le C)$. We further define $\YL = \min(\TL,Y)$ and $\DeltaL = I(\TL \le T, \TL \le C)$. Note that $\DeltaL = 1$ is the at-risk indicator at $\TL$. For subjects who are free of the failure event at $\TL$, one can observe $\TL$ and $\bX(\TL)$ when $\DeltaL = 1$. The training data are $\{ (Y_i,\Delta_i,\YLi,\DeltaL_i,\bX_i(\YLi),\bZ_i), i = 1,\ldots,n \}$, which are assumed to be independent identically distributed replicates of $(Y,\Delta,\YL,\DeltaL,\bX(\YL),\bZ)$.

Define $N(t,a) = \Delta I(Y-a\le t)$. Denote by $\lambda(t\mid a,\bz,\bx)$ the landmark hazard function, that is, $\lambda(t\mid a,\bz,\bx)dt = P(T-\TL \in [t,t+dt) \mid T- \TL\ge t, \TL = a, \bZ = \bz, \bX(\TL) = \bx)$. The survival function $S(t\mid a,\bz,\bx)$ and the hazard function $\lambda(t\mid a,\bz,\bx)$ have a one-to-one correspondence relationship: $S(t\mid a,\bz,\bx) = \exp\{-\int_0^t \lambda(u\mid a,\bz,\bx)du \}$. For $t>0$, we have 
\begin{eqnarray*}
\lefteqn{E\{ N(dt,\YL) - I(Y-\YL\ge t) \lambda(t\mid a,\bz,\bx)dt \mid \DeltaL = 1, \YL=a, \bZ = \bz, \bX(\YL)=\bx  \} }\nonumber\\
&=& E\{ N(dt,a) - I(T\ge a+t, C\ge a+t) \lambda(t\mid a,\bz,\bx)dt \mid C\ge a, T\ge \TL=a, \bZ = \bz, \bX(\TL)=\bx \}\nonumber\\
&=& E\{ I(T-\TL\in [t,t+dt)) - I(T-\TL\ge t) \lambda(t\mid a,\bz,\bx)dt \mid  C\ge a, T\ge \TL=a, \bZ = \bz, \\
&& \bX(\TL)=\bx  \}\times P(C\ge a+t \mid C\ge a, \bZ = \bz)\\
& = & 0.
\end{eqnarray*}
Then we have
\begin{align}
\label{eq:key2}
\lambda(t\mid a,\bz,\bx)dt = \frac{E\{ N(dt,\YL)  \mid \DeltaL = 1, \YL=a,\bZ = \bz, \bX(\YL)=\bx  \}}{E\{  I(Y-\YL\ge t) \mid \DeltaL = 1, \YL=a, \bZ = \bz, \bX(\YL)=\bx  \}}.
\end{align}
Conditioning on $\DeltaL=1$ and $(\YL,\bZ,\bX(\YL))$, subjects with $Y-\YL \ge t$ can be viewed as a representative sample of the population with $T-\TL \ge t$ for each $t>0$. Heuristically, the numerator and denominator in \eqref{eq:key2} can be estimated using partition-based estimators in the subsample with $\DeltaL = 1$. Given a partition $\T$, the function $S_\T(t\mid a,\bz,\bx)$ in \eqref{st} can be estimated by the following estimator,
\begin{align}
\label{Stree}
\nonumber
\hS_{\T}( t \mid a, &\bz,\bx) \\
& = \exp\left\{ -\int_0^{t} \frac{\sumi \DeltaL_i I((\YL_i,\bZ_i, \bX_i(\YL_i))\in l_{\T}(a,\bz,\bx))N_i(du,\YL_i)}{\sumi \DeltaL_i I((\YL_i,\bZ_i,\bX_i(\YL_i))\in l_{\T}(a,\bz,\bx), Y_i-\YL_i \ge u)} \right\}.
\end{align}
When a new subject is event-free at the landmark time $\TL_0$ with predictors $\bZ_0$ and $\bX_0(\TL_0)$, the predicted survival probability based on a single tree is $\hS_\T( t \mid \TL_0, \bZ_0, \bX_0(\TL_0))$. In practice, the partition $\T$ can be constructed via a recursive partition algorithm, and the split-complexity pruning can be applied to determine the size of the tree \citep{leblanc1993survival}. More details of the tree algorithm are given in the Supplementary Material \citep{supp}.

\subsection{Survival tree ensembles based on martingale estimating equations}
\label{sect:ensemble}

Since the prediction based on a tree is often unstable, ensemble methods such as bagging \citep{breiman1996bagging} and random forests \citep{breiman2001random} have been commonly applied. The original random forests perform the prediction for a new data point by averaging predictions from a large number of trees, which are often grown sufficiently deep to achieve low bias \citep{friedman2001elements}. However, for censored data, a large tree may result in a small number of observed failures in the terminal nodes, leading to increased estimation bias of survival or cumulative hazard functions. Existing survival forests inherit from the original random forest and directly average the cumulative hazard prediction from individual trees. Therefore, the node size parameter needs to be carefully tuned to achieve accurate prediction. On the other hand, if the target estimate can be expressed as the solution of an unbiased estimating equation, a natural way is to solve the averaged estimating equations. In what follows, we propose an ensemble procedure based on averaging martingale estimating equations.

For $b = 1,\ldots,B$, we draw the $b$th bootstrap sample from the training data. Let $\mathbb{T} = \{\T_b\}_{b = 1}^B$ be a collection of $B$ partitions constructed using the bootstrap datasets. Each partition is constructed via a recursive partition procedure where at each split, $m$ predictors are randomly selected as the candidate variables for splitting, and $m$ is smaller than the number of predictors. Let $l_b$ be the partition function based on the partition $\T_b$. The tree-based estimation from $\T_b$ can be obtained from the following estimating equation,
\begin{align*}
\sumi w_{bi} I( (\YL_i,\bZ_i,\bX_i(\YL_i))\in l_b(a,\bz,\bx) ) \DeltaL_i 
\{ N_i(dt,\YL_i) - I(Y_i-\YL_i\ge t) \lambda(t\mid a,\bz,\bx) dt\} = 0,
\end{align*}
where $w_{bi}$ is the frequency of the $i$th observation in the $b$th bootstrap sample. Note that when $w_{bi}=1$ for all $i=1,\ldots,n$, solving the above estimating equation yields the estimator in \eqref{Stree}. To perform prediction using all the trees, we consider the following averaged estimating equation,
\begin{align*}
\sumi  w_i(a,\bz,\bx)
\{ N_i(dt,\YL_i) - I(Y_i-\YL_i\ge t) \lambda(t\mid a,\bz,\bx) dt\} = 0,
\end{align*}
where 
$w_i(a,\bz,\bx) = \sumb w_{bi}  I\{ (\YL_i,\bZ_i,\bX_i(\YL_i))\in l_{b}(a,\bz,\bx) \}\DeltaL_i/B$. Solving the averaged estimating equation yields
\begin{align*}
\hS_{\mathbb{T}}( t \mid a, \bz, \bx) = \exp\left\{ -\int_0^{t} \frac{\sumi w_i(a,\bz,\bx) N_i(ds,\YL_i)}{\sumi w_i(a,\bz,\bx) I( Y_i-\YL_i \ge s)} \right\}.
\end{align*}
The estimator $\hS_{\mathbb{T}}( t \mid a, \bz, \bx)$ can be viewed as an adaptive nearest neighbour estimator \citep{lin2006random}, where the weight assigned to each observation comes from random forests. The risk prediction algorithm is given in Table \ref{tab:a1}.
{
	\spacingset{1}
\begin{table}[th]
\footnotesize
	\caption{The data preprocessing and risk prediction algorithm}
	\label{tab:a1}
\begin{algorithm}[H]
\SetAlgoLined
\SetKwInOut{Input}{Input}
\SetKwInOut{Output}{Output}

		\caption{The survival tree ensemble algorithm}
		\Input{$\{ (Y_i,\Delta_i,\YLi,\DeltaL_i,\bX_i(\YLi),\bZ_i), i = 1,\ldots,n \}$ }
		\Output{
		 $\widehat S_{\mathbb{T}}(t\mid a,\bz, \bx)$}

		\smallskip

		\SetKwFunction{FMain}{Transform}
    \SetKwProg{Fn}{Function}{:}{end}
    
  \Fn{\FMain{$a,\bx$}}{ {\tiny\tcc{The function transforms the longitudinal predictors $\bX(\TL)$ (i.e., $\bx$) to a vector $\widetilde\bX(\TL)$ (i.e., $\widetilde\bx$) such that the rules in (R1) and (R2) on $\bX(\TL)$ can be achieved using a conventional tree-structured rule on $\widetilde\bX(\TL)$}}
    $\widetilde\bx \leftarrow {\rm vector}( {\rm size}: 2Kq + J)$ \;
    
        \For{ $l = 1$ {to} $Kq + J$}
        {
             \uIf{$ 1 \le l \le Kq $ and $\bx[l] \neq {\rm NA}$
             \tiny\tcp*{observed longitudinal measurements}
             }{
      $\widetilde\bx[2l-1] \leftarrow \bx[l]$ \; 
      $\widetilde\bx[2l] \leftarrow \bx[l]$ \;
  }
  \uElseIf{$ 1 \le l \le Kq $ { and } $\bx[l] = {\rm NA}$
  \tiny\tcp*{unobserved longitudinal measurements}}{
      $\widetilde\bx[2l-1] \leftarrow M$ \;
      $\widetilde\bx[2l] \leftarrow -M$ \;
  }
  {\tiny\tcc{The split $\widetilde\bx[2l-1]>c$ vs. $\widetilde\bx[2l-1]\le c$ corresponds to ($\bx[l] > c$ or $\bx[l] = {\rm NA}$) vs. $\bx[l]\le c$;
  the split $\widetilde\bx[2l]>c$ vs. $\widetilde\bx[2l]\le c$ corresponds to $\bx[l] > c$ vs. ($\bx[l]\le c$ or $\bx[l] = {\rm NA}$)}}
  \uElseIf{$ Kq + 1 \le l \le Kq + J $ and $\bx[l] = a+$
  \tiny\tcp*{unobserved intermediate events}}{
      $\widetilde\bx[Kq + l] \leftarrow M$ \;
  }
  \Else{
    $\widetilde\bx[Kq + l] \leftarrow \bx[l]/a$ 
    \tiny\tcp*{observed intermediate events}
  }
{\tiny\tcc{The split $\widetilde\bx[Kq+l]>c$ vs. $\widetilde\bx[Kq+l]\le c$ corresponds to $\bx[l]/a > c$ vs. $\bx[l]/a\le c$}}
        }

        \KwRet $\widetilde\bx$ \;}

\smallskip
		\textbf{Data Preprocessing}\\

		\For{$i = 1$ to $n$}
		{
				\If{$\DeltaL_i = 1$}
				{
				$ \widetilde\bX_i(\YL_i) \leftarrow$ \FMain{$\YL_i, \bX_i(\YL_i)$}\;
				}
		}
       $\widetilde\bx \leftarrow$ \FMain{$a, \bx$} \;

       \smallskip
		\textbf{Survival probability prediction}
		
		\noindent {Initialize the weights: 
		$w_i \leftarrow 0$, $i = 1,\ldots, n$\;}

		\For{$b = 1$ to $B$}{
			Draw the $b$th bootstrap sample from the training data\;

			 Construct a partition $\T_b$ using subjects with $\Delta_L = 1$ in the $b$th bootstrap sample: 
			 \begin{itemize}
			     \item The predictors are $\{\YL,\bZ,\widetilde\bX(\YL)\}$ and the censored outcome is $(Y-\YL, \Delta)$\;
			     \item At each split, a random selection of $m$ predictors are used as the candidate splitting variables\;
			 \end{itemize}

		\For{$i = 1$ to $n$}
		{
				\If{ $\DeltaL_i = 1$ and $(\YL_i, \bZ_i,\widetilde\bX_i(\YL_i))$ and $(a,\bz,\widetilde\bx)$ are in the same terminal node}
				{
				$w_{i} \leftarrow w_{i} +  w_{bi}$\;
				{\tiny\tcc{$w_{bi}$ is the frequency of the $i$th observation in the $b$th bootstrapped sample}}
				}
		}
		
	}		
			
		 Predict the survival probability using 
		$$\widehat S_{\mathbb{T}}(t\mid a,\bz, \bx) = \exp\left\{-\sum_{i=1}^{n} \frac{w_i\Delta_iI(Y_i -\YL_i \le t) }{ \sum_{j=1}^n w_{j}I(Y_j-\YL_j \ge Y_i-\YL_i)}\right\}.$$

\end{algorithm}
\end{table}
}

\section{Evaluating the landmark prediction performance}
\label{sect:con}

To evaluate the performance of the predicted risk score, we extend the cumulative/dynamic receiver operating characteristics (ROC) curves \citep{heagerty2000time}, which has been commonly used when a risk score is based on baseline predictors.
We note that ROC and concordance indices for dynamic prediction at a fixed landmark time has been studied in the literature \citep{rizopoulos2017dynamic,wang2017dynamic}. Here we consider a more general case where the landmark time $\TL$ is random and subject to right censoring. For $t>0$, subjects with $0\le T-\TL< t$ are considered as cases and subjects with $T-\TL\ge t$ are considered as controls. The ROC curve then evaluates the performance of a risk score that discriminates between subjects who have experienced the events prior to $\TL+t$ and those who do not. 

Let $g(t,\TL,\bZ,\bX(\TL))$ denote a risk score based on $(\TL,\bZ,\bX(\TL))$, with a larger value indicating a higher chance of being a case. For each $t>0$, The true positive rate and false positive rate at a threshold of $c$ are defined as follows,
\begin{align*}
{\TPR_t(c)} =&  P( g(t,\TL,\bZ,\bX(\TL)) > c \mid 0\le T-\TL< t, {\TL \le \tau_0}),\\
{\FPR_t(c)} =&  P( g(t,\TL,\bZ,\bX(\TL)) > c \mid  T-\TL \ge t, {\TL \le \tau_0}),
\end{align*}
{where $\tau_0$ is a pre-specified constant.} The ROC curve at $t$ is defined as ${\rm ROC}_t(p) = \TPR_t(\FPR_t^{-1}(p))$. Following the arguments of \cite{mcintosh2002combining}, it can be shown that $g(t,a,\bz,\bx) = 1 - S(t\mid a,\bz,\bx )$ yields the highest ROC curve, which justifies the use of the proposed time-dependent ROC curve.
Moreover, the area under the ROC curve is equivalent to the following concordance measure, 
\begin{eqnarray*}
 \lefteqn{\CON_t(g)}  \\
&=& P\{g(t,\TL_1,\bZ,\bX_1(\TL_1)) < g(t,\TL_2,\bZ,\bX_2(\TL_2)) \mid \\
&& \hspace{2.2in} 0\le T_2-\TL_2 < t \le T_1-\TL_1, {\TL_1 \le \tau_0, \TL_2 \le \tau_0}\} +\\
&& \hspace{0.01in} 0.5 P\{ g(t,\TL_1,\bZ,\bX_1(\TL_1)) = g(t,\TL_2,\bZ,\bX_2(\TL_2)) \mid \\
&& \hspace{2.2in} 0\le T_2-\TL_2 < t \le T_1-\TL_1, {\TL_1 \le \tau_0, \TL_2 \le \tau_0}\},
\end{eqnarray*}
where $(\TL_1,\bZ_1,\bX_1(\TL_1),T_1)$ and $(\TL_2,\bZ_2,\bX_2(\TL_2),T_2)$  are independent pairs of observations, and the second term accounts for potential ties in the risk score. 

In practice, one usually builds the model on a training dataset and evaluates its performance on an independent test dataset that are also subject to right censoring. To simplify notation here, we construct the estimator for $\CON_t(g)$ using the observed data introduced in Section \ref{sect:censor}, although $\CON_t$ evaluated using the test data should be used in real applications. 
Define $ d_{ij}(t)= g(t,\YL_j,\bZ_j,\bX_j(\YL_j) )\bl{-}g(t, \YL_i, \bZ_i, \bX_i(\YL_i))$. The $\CON_t(g)$ measure can be estimated by
\begin{eqnarray}
\lefteqn{\widehat{\CON}_t(g)} \label{con1}\\ 
&=& \frac{\sum_{i\neq j}  \{I(d_{ij}(t)>0) + 0.5I(d_{ij}(t)=0) \}\frac{\Delta_jI( 0\le Y_j-\YL_j \le t < Y_i-\YL_i, {\YL_i \le \tau_0, \YL_j \le \tau_0})}{\hS_C(Y_j\mid \bZ_j)\hS_C(\YL_i+t\mid \bZ_i) }   }{\sum_{i\neq j} \frac{\Delta_jI( 0\le Y_j-\YL_j \le t < Y_i-\YL_i, {\YL_i \le \tau_0, \YL_j \le \tau_0})}{\hS_C(Y_j\mid \bZ_j)\hS_C(\YL_i+t\mid \bZ_i) }   } ,
\nonumber
\end{eqnarray}
where $\widehat S_C(t\mid \bz)$ is an estimator for the conditional censoring distribution $S_C(t\mid \bz) = P(C\ge t\mid \bZ = \bz)$. For example, survival trees or forests can be applied to estimate $S_C(t\mid \bz)$; when censoring is completely random, the Kaplan-Meier estimator can also be applied. Under regularity conditions, we show that $\widehat{\CON}_t(g)$ consistently estimates ${\CON}_t(g)$ in the Supplementary Material.
In practice, one can either use the concordance at a given time point $t$ or an integrated measure to evaluate the overall concordance on a given time interval $[t_{\rm L},t_{\rm U}]$. In the latter case, a weighted average of concordance on a grid of time points in $[t_{\rm L},t_{\rm U}]$ can be reported. For example, one may assign equal weights to all the time points. Another example is a weight proportional to the denominator in $\widehat{\rm CON}_t(g)$, that is, $\sum_{i\neq j} {\Delta_j I( 0\le Y_j-\YL_j \le t < Y_i-\YL_i, {\YL_i \le \tau_0, \YL_j \le \tau_0})}/{\hS_C(Y_j\mid \bZ_j)\hS_C(\YL_i+t\mid \bZ_i) } $, to avoid potentially unstable estimation for very small or large time points.

\section{Permutation variable importance}
\label{sect:vimp}

Variable importance is a useful measure for understanding the impact of predictors in tree ensembles and can be used as a reference for variable selection \citep{breiman2001random}. In the original random forests, each tree is constructed using a bootstrap sample of the original data, and the out-of-bag (OOB) data can be used to estimate the OOB prediction performance. The permutation variable importance of a predictor is computed as the average decrease in model accuracy on the OOB samples when the respective feature values are randomly permuted.

To study variable importance in dynamic risk prediction using censored data, we consider an extension of variable importance. Following the original random forests, the OOB prediction for each training observation is made based on trees constructed without using this observation. Applying the same arguments as in Section \ref{sect:con}, the prediction based on trees built without the $i$th subject is
\begin{align*}
\hS_{-i}(t\mid a,\bz,\bx) = \exp\left\{ -\int_0^t \frac{\sumk w_{k,-i}(a,\bz,\bx)N_k(ds,\YL_k)}{\sumk w_{k,-i}(a,\bz,\bx)I(Y_k-\YL_k\ge s)} \right\},
\end{align*}
where {$w_{k,-i}(a,\bz,\bx) = \sumb w_{bk} I( (\YL_k,\bZ_k,\bX_k(\YL_k))\in l_b(a,\bz,\bx)  , w_{bi} = 0)\DeltaL_i$.} Define $d_{ij}(t) = - \hS_{-i}(t\mid \YL_i,\bZ_i,\bX_i(\YL_i))+\hS_{-j}(t\mid \YL_j,\bZ_j,\bX_j(\YL_j))$. The OOB concordance at $t$ can be calculated by applying \eqref{con1}. To compute variable importance for a predictor, we permute this predictor and calculate the OOB concordance after permutation. We repeat the permutation multiple times (e.g., 100 times) and define the variable importance as the average difference in OOB concordances over all the permutations.

Permuting variables is straightforward in the case of fixed landmark times (i.e., $\TL = t_k$), where we randomly shuffling the observed values of the predictor among {individuals who remained under observation at the landmark time, that is, subjects with $\DeltaL = 1$.} When the landmark time is random (i.e., $\TL = U_j$), we propose different permutation procedures for calculating variable importance according to the type of the variable: (1) If the variable of interest is completely observed at the landmark time $\TL$ (e.g., the value of $\TL$ and baseline covariates $\bZ$), we randomly shuffle its values among subjects with $\DeltaL = 1$. (2) If the variable of interest is an intermediate event time $U_{j'} (j' \neq j)$, we propose to permute its relative value to the landmark time, that is,  $U_{j'}(\TL)/\TL$, among subjects with $\DeltaL = 1$. Note that the reason to permute the ratio $U_{j'}(\TL)/\TL$ instead of the untransformed intermediate event time directly is to avoid incompatible pairs of the intermediate event time and the landmark time under permutation. (3) If the variable of interest is an element of $\bW(t_k)$, we permute its values among subjects with complete observations at the landmark time (i.e., $\TL \ge t_k$ and $\DeltaL = 1$). This is because $\bW(t_k)$ is not used in prediction for subjects with $\TL < t_k$.

\section{Simulation}
\label{sect:simu}

Simulation studies were conducted to assess the performance of the proposed methods in estimating the landmark survival probability in \eqref{pr}
under scenarios when the landmark time is fixed or random. 
In both cases, we generated the time-independent predictors $\bZ=(Z_1, \ldots, Z_{10})$ from a multivariate normal distribution
with $E(Z_i) = 1$, ${\rm Var}(Z_i) = 1$, and ${\rm Cov}(Z_i, Z_j) = 0$, for $i, j = 1, \ldots, 10$. 
The longitudinal predictors were observed intermittently at time points specified below.

In the first set of simulations, 
the longitudinal predictors $\bW(\cdot)$ were measured at fixed landmark time points $t_k = k$ for $k=1,\ldots,K$.
The longitudinal predictors $\bW(t) = (W_1(t), \ldots, W_{10}(t))$ were generated from 
$W_i(t) = a_iF(b_i t) / t$, $i = 1, \ldots, 10$, 
where $a_i$ and $b_i$ are independent standard uniform random variables and $F(x) = 1 - \exp(-x^2)$. 
The probability in \eqref{pr} can be expressed as
$P(T \ge t_{k} + t \mid T\ge t_{k}, \bW(t_1),\ldots ,\bW(t_{k}), \bZ)$
for $0 = t_0 < t < t_{k+1}-t_k$, $k\ge 1$, and $P(T \ge  t \mid \bZ)$ for $0<t<t_1$. 
For $k\ge 1$, we assume the hazard function of $T$ on $(t_k,t_{k+1})$ depends on the history of $\bW(\cdot)$ up to $t$ only through its value at $t_k$. 
For $t\in (t_k,t_{k+1})$ and $k\ge 0$, we consider the following hazard functions for $T$,\\
(I) $\lambda(t \mid \mH_W(t), \bZ ) =
t^2\exp\left\{-5 +  \sum_{j=1}^{10} \alpha_{kj}W_j(t_k) + \sum_{j=1}^{10} \beta_{kj} W_j(t_k)Z_j + \sum_{j = 1}^3 Z_j^2\right\}$, $\alpha_{kj} = \beta_{kj} = 2I(k=1)+4I(k\ge 2)$ for $1\le j\le 3$, and $\alpha_{kj} = \beta_{kj} = 0$ for $4\le j\le 10$;\\
(II) $\lambda(t \mid \mH_W(t), \bZ ) =
0.1t^2 + \exp\left\{-5 +  \sum_{j=1}^{10} \alpha_{kj}W_j(t_k) + \sum_{j=1}^{10} \beta_{kj} W_j(t_k)Z_j + \sum_{j = 1}^3 Z_j^2\right\}$, $\alpha_{kj} = \beta_{kj} = I(k=1)+2I(k\ge 2)$ for $1\le j\le 3$, and $\alpha_{kj} = \beta_{kj} = 0$ for $4\le j\le 10$.\\
The closed-form expressions of the true landmark survival probabilities under Models (I) and (II) are given in the Supplementary Material. When evaluating the prediction performance of different approaches, we focus on the landmark probability at $t_2 = 2$.

In the second set of simulations, we consider the case where both an intermediate event and longitudinal markers are present. Specifically, we generated the event times from irreversible multi-state models with three states: healthy, diseased, and death. 
We assume that all subjects started in the healthy state, disease onset is an intermediate event, and death is the event of interest.
We generated the time to the first event, denoted by $D$, from a uniform distribution on [0, 5].
Define the disease indicator, $\Pi$, 
where $\Pi = 1$ indicates the subject moves from the healthy state to the disease state at time $D$, 
and $\Pi= 0$ indicates the subject moves from the healthy state to death at time $D$.
The disease indicator was obtained via a logistic regression model,
$\logit P(\Pi=1\mid \bZ, \bW(D)) = \sum_{j = 1}^3 W_i(D) + \sum_{j=1}^3Z_j + \gamma$, where
$\gamma$ is a frailty variable following a gamma distribution with mean 1 and variance 0.5,
$\bW(t) = (W_1(t), \ldots, W_{10}(t))$ were generated from 
$W_j(t) = a_j \{1 - \exp(-0.04t^2)\}$, and $a_j$ follows a uniform distribution on $[-1, 1]$ for $j = 1,\ldots,10$.
Given a subject had developed the disease at time $D$, i.e., $\Pi = 1$, the residual survival time, $R$, 
was generated from the following models,\\
(III) $\log R = -5 + \sum_{j = 1}^3W_j(D) + \sum_{j = 1}^3Z_j^2 + \sum_{j = 1}^3 W_j(D) Z_{j} + \log(1 + D) + \gamma + \epsilon$,
where $\epsilon$ is a standard normal random variable;\\
(IV) $\log R = -5 + \sum_{j = 1}^3W_j(D) + \sum_{j = 1}^3Z_j^2 + \sum_{j = 1}^3 W_j(D) Z_{j} + \log(1 + D) + \gamma + 
I(Z_1 > 0)\epsilon_1 + I(Z_1 \le 0)\epsilon_2$,
where $\epsilon_1$ and $\epsilon_2$ are independent normal random variables with variances 1 and 0.25, respectively;\\
(V) The hazard function of $R$ is
$t^2\exp [-5 +  \sum_{j=1}^{3} \{2I(1\le D < 2)+4I(D\ge 2)\}\{W_j(D) + W_j(D)Z_j + Z_j^2\} ].$\\
When $\Pi = 1$, the time to death is $T = D+R$, and the time to the intermediate event is $U = D$; 
when $\Pi = 0$, the time to death is $T = D$ and the intermediate event does not occur.
Under Models (III)-(V), we consider the following three scenarios depending on how the landmark time and the longitudinal markers are observed:
\begin{enumerate}
\item[(A)] The landmark is the intermediate event, i.e., $\TL = U$, and $\bW(\cdot)$ is observed intermittently at $t_k = k$, {$k = 1,\ldots,5$}. The target probability in \eqref{pr} is
$$P(T\ge  U +t \mid T\ge  U, U, \bW(t_1,U), \ldots, \bW(t_K, U), \bZ).$$
\item[(B)] The landmark time is fixed at $\TL = a$, and $\bW(\cdot)$ is observed at the intermediate event. The target probability is 
\begin{align*}
\begin{cases}
    P( T-a\ge t\mid T\ge a, U ,\bW(U) , \bZ ), & U\le a,\\
   P( T-a\ge t\mid T\ge a, U > a,\bZ  ), & U>a.\\ 
\end{cases}
\end{align*}
\item[(C)] The landmark is the intermediate event, and $\bW(\cdot)$ is observed at the intermediate event. The target probability is $P(T\ge  U +t \mid T\ge  U, U, \bW(U), \bZ )$.
\end{enumerate}
Scenario (A) is motivated by the CFFPR data, where the longitudinal marker is regularly monitored. Scenarios (B) and (C) are motivated by applications where markers are observed when a disease is diagnosed. In Scenario (B), we set $a=2$. Due to the complicated relationship between event times and longitudinal markers, deriving the closed-form expression of the true probability under Models (III)-(V) is challenging.
We outline the the Monte Carlo method used to approximate the true probabilities in the Supplementary Material.

For all scenarios, the censoring time was generated from an independent exponential distribution with rate $c$,
where $c$ was chosen to achieve either a 20\% or 40\% rate of censoring at the baseline. 
We simulated 500 training datasets with sample sizes of 200 and 400 at baseline. The results for large sample sizes ($n=5000$) are included in the Supplementary Material.
The trees were constructed with a minimum terminal node size of 15. 
When a single tree was used for prediction, the size of the tree was determined by split-complexity pruning via ten-fold cross-validation. 
To grow the trees in the ensemble method, we randomly selected 
$\left \lceil \sqrt p \right \rceil$ variables at each splitting step and did not prune the trees. We applied the log-rank splitting rule in the ensemble method in order to compare the martingale-based ensemble approach with the default ensemble approach in random survival forests.
Each fitted model was evaluated on independent test data with 500 observations. 
The evaluating criteria were the integrated mean absolute error (IMAE), the integrated mean squared error (IMSE), the integrated Brier score (IBS), and the integrated concordance over $[0,t_0]$, 
where $t_0$ is set to be approximately the 90\% quantile of $T-\TL$.
The integrated concordance was defined in Section 4, and other evaluating criteria are defined as follows:
\begin{align*}
{\rm IMAE} &= \sum_{i=1}^{500} \int_0^{t_0} \left| \widehat S(t\mid T^0_{{\rm L}i},\bZ_i^0,\mH_i^0(T^0_{{\rm L}i})) - S(t\mid T^0_{{\rm L}i},\bZ_i^0,\mH_i^0(T^0_{{\rm L}i}))\right| dt/500,\\
{\rm IMSE} &= \sum_{i=1}^{500} \int_0^{t_0} \left\{ \widehat S(t\mid T^0_{{\rm L}i},\bZ_i^0,\mH_i^0(T^0_{{\rm L}i})) - S(t\mid T^0_{{\rm L}i},\bZ_i^0,\mH_i^0(T^0_{{\rm L}i}))\right\}^2 dt/500,\\
{\rm IBS} &= \sum_{i=1}^{500} \int_0^{t_0} \left\{ \widehat S(t\mid T^0_{{\rm L}i},\bZ_i^0,\mH_i^0(T^0_{{\rm L}i})) - I(T^0_i-\TL^0_i \ge t)\right\}^2 dt/500,
\end{align*}
where  
$S(t\mid  \TL , \bZ , \mH(\TL))= P(T-\TL \ge t\mid  T\ge \TL,  \TL , \bZ, \mH(\TL))$ is the target survival probability, and the superscript 0 is used to denote the test data.

For comparison, we applied the conventional random survival forest implemented in the R package \emph{ranger} \citep{ranger} and the simple landmark Cox model in all scenarios; we also applied the two-stage landmark approach in Models (I) and (II).
We used the proposed data preprocessing procedure to prepare the predictors for the conventional random survival forest but calculated the predicted survival probabilities by default (i.e., averaging the cumulative hazard predictions).
Under Models (I) and (II), the predictors in the simple landmark Cox model are $\{\bW(t_1),\ldots,\bW(t_k),\bZ \}$.
Under Models (III)-(V), the predictors in the simple landmark Cox model are
$\{ U, \bZ, \bW(t_k)I(t_k\le U), I(t_k>U); k\ge 1 \}$, $\{ \bZ, UI(U \le a ),\bW(U)I(U \le a ), I(a>U) \}$, and 
$\{ U, \bZ, \bW(U)\}$ in sub-scenarios (A), (B), and (C), respectively.
For the two-stage landmark approach, we fit separate linear mixed models for all the variables in $\bW(\cdot)$: each model includes a time variable and a random intercept, and were estimated using repeated measurements from subjects who are at-risk at the landmark time. The predictors of the landmark Cox model then include the BLUPs of random effects and $\bZ$.

The simulation results are summarized in Tables~\ref{sim} and \ref{sim2}, in which
the proposed ensemble method outperforms the others based on the four evaluation criteria we considered.
As expected, when the sample size increases from 200 to 400, the IMAEs, the IMSEs, and the IBSs of the proposed methods decrease, while the integrated concordance increases.
The conventional random survival forest approach performs similarly to the proposed ensemble method when $n = 200$ but loses its edge as the sample size increases.
On the other hand, the simple landmark Cox model and the two-stage landmark approach yield similar results under Model (I), but the former yields better results under Model (II).
When comparing the tree models with the Cox models,
we observe that a small IMSE does not necessarily accompany by a large integrated concordance. 
We conjecture this is because the concordance measure only depends on the order of the predicted survival probability and is less sensitive in terms of risk calibration.
In summary, the proposed ensemble method has strong performances and serves as an appealing tool for dynamic risk prediction.

\begin{table}[h!]
  \centering
\caption{Summaries of integrated mean absolute error (IMAE) $(\times1000)$, integrated mean squared error (IMSE) $(\times1000)$, integrated Brier score (IBS) $(\times1000)$, 
 and integrated concordance (ICON) $(\times1000)$ of different methods in the first set of simulations. 
 cen is the censoring percentage;
 Tr: the proposed tree; E1: the proposed survival tree ensemble; E2: the original random survival forest;
 C1: the landmark Cox model; C2: the two stages landmark Cox model.}
  \setlength{\tabcolsep}{2.6pt}
  \label{sim}
  \begin{tabular}{ll ccccc ccccc ccccc ccccc}
    \toprule 
    && \multicolumn{5}{c}{IMAE} & \multicolumn{5}{c}{IMSE} & \multicolumn{5}{c}{IBS} & \multicolumn{5}{c}{ICON} \\
    \cmidrule(l){3-7}\cmidrule(l){8-12}\cmidrule(l){13-17}\cmidrule(l){18-22}
    $n$  & cen & Tr & E1 & E2 & C1 & C2 & Tr & E1 & E2 & C1 & C2 & Tr & E1 & E2 & C1 & C2 & Tr & E1 & E2 & C1 & C2 \\
    \midrule
    &&\multicolumn{20}{c}{Scenario (I)}\\
    200 & 20\% &   213 & 196 & 219 & 220 & 216 & 74 & 64 & 71 & 91 & 91 &   192 & 183 & 196 & 208 & 208 & 508 & 642 & 634 & 572 & 624 \\ 
    & 40\% &    215 & 202 & 222 & 236 & 239 & 76 & 68 & 74 & 106 & 112 & 185 & 177 & 188 & 212 & 217 & 509 & 621 & 615 & 562 & 607 \\
    400 & 20\% &   201 & 172 & 207 & 204 & 197 & 65 & 48 & 63 & 74 & 66 &   181 & 165 & 188 & 192 & 184 & 515 & 694 & 687 & 593 & 655 \\ 
    & 40\% &   205 & 178 & 212 & 209 & 206 & 70 & 51 & 66 & 80 & 73 &   176 & 160 & 180 & 188 & 181 & 526 & 675 & 666 & 586 & 647 \\
    [1ex]
    &&\multicolumn{20}{c}{Scenario (II)}\\
    200 & 20\% &  73 & 76 & 83 & 139 & 175 & 13 & 13 & 14 & 38 & 58 & 172 & 172 & 182 & 196 & 214 & 531 & 537 & 537 & 533 & 534 \\
    & 40\% &            76 & 79 & 86 & 192 & 236 & 13 & 14 & 15 & 69 & 98 & 161 & 161 & 171 & 208 & 235 & 530 & 538 & 536 & 536 & 536 \\
    400 & 20\% &       64 & 59 & 81 &  91 & 114 & 11 & 10 & 13 & 19 & 26 & 164 & 160 & 172 & 177 & 184 & 530 & 547 & 546 & 539 & 541 \\ 
    & 40\% &           62 & 64 & 83 & 113 & 145 & 10 & 11 & 14 & 26 & 41 & 156 & 150 & 159 & 172 & 184 & 531 & 544 & 542 & 536 & 538 \\
    \bottomrule
  \end{tabular}
\end{table}

\begin{table}[h!]
	\spacingset{1}
  \centering
\caption{Summaries of integrated mean absolute error (IMAE) $(\times1000)$, integrated mean squared error (IMSE) $(\times1000)$, integrated Brier score (IBS) $(\times1000)$, 
 and integrated concordance (ICON) $(\times1000)$ of different methods in the second set of simulations. 
 cen is the censoring percentage;
 Tr: the proposed tree; E1: the proposed survival tree ensemble; E2: the original random survival forest;
 C1: the landmark Cox model.}
  \setlength{\tabcolsep}{4.0pt}
  \label{sim2}
    	\footnotesize
  \begin{tabular}{cc cccc cccc cccc cccc}
    \toprule 
    && \multicolumn{4}{c}{IMAE} & \multicolumn{4}{c}{IMSE} & \multicolumn{4}{c}{IBS} & \multicolumn{4}{c}{ICON} \\
    \cmidrule(l){3-6}\cmidrule(l){7-10}\cmidrule(l){11-14}\cmidrule(l){15-18}
     $n$  & cen & Tr & E1 & E2 & C1 & Tr & E1 & E2 & C1 & Tr & E1 & E2 & C1 & Tr & E1 & E2 & C1 \\
    \midrule
    &&\multicolumn{16}{c}{Scenario (III-A)}\\
     200 & 20\% &  192 & 167 & 178 & 310 & 84 &  72 & 79 & 237 &  98 & 79 & 86 & 260 & 816 & 902 & 900 & 531 \\ 
    & 40\% &               266 & 194 & 198 & 405 & 114 & 85 & 93 & 319 & 140 & 81 & 87 & 323 & 610 & 906 & 906 & 530 \\ 
     400 & 20\% &          173 & 154 & 169 & 255 & 68 &  61 & 69 & 185 &  70 & 67 & 78 & 210 & 904 & 909 & 909 & 544 \\ 
    & 40\% &               204 & 178 & 191 & 323 & 85 &  76 & 84 & 239 &  74 & 64 & 76 & 241 & 882 & 917 & 915 & 523 \\ 
    &&\multicolumn{16}{c}{Scenario (III-B)}\\
    200 & 20\% &   234 & 216 & 219 & 249 & 94 &  81 & 89 & 122 & 143 & 132 & 133 & 161 & 544 & 713 & 700 & 614 \\
    & 40\% &               245 & 228 & 215 & 268 & 99 &  83 & 87 & 138 & 144 & 133 & 133 & 169 & 545 & 652 & 645 & 584 \\
     400 & 20\% &          212 & 181 & 192 & 213 & 84 &  65 & 70 & 92 &  132 & 116 & 126 & 133 & 572 & 764 & 760 & 646 \\
    & 40\% &               224 & 196 & 193 & 226 & 92 &  67 & 68 & 101 & 135 & 117 & 127 & 136 & 584 & 708 & 703 & 619 \\
    &&\multicolumn{16}{c}{Scenario (III-C)}\\
    200 & 20\% &   236 & 214 & 228 & 259 & 108 &  93 & 97 &  125 & 150 & 138 & 138 & 168 & 524 & 822 & 713 & 643 \\
    & 40\% &               253 & 233 & 256 & 292 & 121 & 106 & 110 & 150 & 160 & 147 & 147 & 187 & 522 & 783 & 678 & 623 \\
     400 & 20\% &          232 & 187 & 206 & 236 &  95 &  66 & 85 &   98 & 138 & 113 & 128 & 143 & 550 & 901 & 810 & 688 \\
    & 40\% &               247 & 204 & 233 & 258 & 106 &  77 & 94 &  114 & 146 & 122 & 134 & 153 & 574 & 872 & 777 & 669 \\
    &&\multicolumn{16}{c}{Scenario (IV-A)}\\
    200 & 20\% &   183 & 163 & 170 & 311 &  80 & 69 & 69 & 238 &  94 & 80 & 81 & 261 & 844 & 902 & 876 & 519 \\ 
    & 40\% &              248 & 191 & 195 & 411 & 109 & 83 & 81 & 324 & 127 & 81 & 83 & 329 & 680 & 906 & 871 & 526 \\ 
     400 & 20\% &         168 & 150 & 165 & 258 &  74 & 59 & 66 & 187 &  79 & 67 & 78 & 209 & 906 & 909 & 879 & 510 \\ 
    & 40\% &              197 & 173 & 187 & 325 &  90 & 72 & 79 & 241 &  83 & 64 & 76 & 240 & 885 & 915 & 876 & 510 \\ 
    &&\multicolumn{16}{c}{Scenario (IV-B)}\\
    200 & 20\% &   199 & 180 & 181 & 206 & 73 & 62 & 63 & 91 & 142 & 132 & 134 & 161 & 543 & 717 & 706 & 615 \\
    & 40\% &              198 & 178 & 177 & 216 & 72 & 59 & 60 & 98 & 144 & 133 & 133 & 168 & 549 & 655 & 648 & 585 \\
     400 & 20\% &         184 & 153 & 156 & 190 & 65 & 49 & 53 & 72 & 130 & 115 & 125 & 132 & 581 & 770 & 770 & 647 \\
    & 40\% &              186 & 152 & 157 & 188 & 66 & 46 & 50 & 72 & 134 & 117 & 127 & 135 & 580 & 712 & 707 & 620 \\
    &&\multicolumn{16}{c}{Scenario (IV-C)}\\
    200 & 20\% &   269 & 245 & 243 & 263 & 114 &  98 & 103 & 130 & 150 & 136 & 137 & 167 & 520 & 828 & 833 & 649 \\
    & 40\% &              299 & 276 & 275 & 297 & 128 & 112 & 117 & 155 & 159 & 146 & 146 & 185 & 522 & 789 & 789 & 631 \\
     400 & 20\% &         252 & 209 & 212 & 244 & 101 &  81 & 90 &  114 & 147 & 111 & 125 & 151 & 552 & 904 & 897 & 694 \\
    & 40\% &              278 & 239 & 240 & 274 & 114 &  93 & 100 & 131 & 155 & 120 & 132 & 162 & 571 & 878 & 871 & 674 \\
    &&\multicolumn{16}{c}{Scenario (V-A)}\\
    200 & 20\% &  212 & 185 & 190 & 349 &  89 & 75 & 78 & 311 & 133 & 112 & 115 & 333 & 688 & 832 & 824 & 534 \\ 
    & 40\%  &           229 & 208 & 215 & 368 & 103 & 90 & 92 & 350 & 115 & 100 & 103 & 316 & 707 & 815 & 820 & 526 \\ 
     400 & 20\% &       193 & 166 & 179 & 377 &  75 & 66 & 69 & 310 & 107 & 105 & 105 & 362 & 828 & 859 & 843 & 534 \\ 
    & 40\% &            220 & 190 & 206 & 427 &  91 & 83 & 84 & 365 &  95 & 92 & 93 & 347 & 810 & 847 & 842 & 539 \\
    &&\multicolumn{16}{c}{Scenario (V-B)}\\
    200 & 20\% &  180 & 158 & 154 & 188 & 54 & 42 & 42 & 70 & 161 & 150 & 151 & 175 & 564 & 672 & 677 & 572 \\ 
    & 40\%  &           189 & 164 & 164 & 200 & 62 & 46 & 47 & 83 & 147 & 135 & 138 & 164 & 569 & 650 & 651 & 566 \\ 
     400 & 20\% &       168 & 136 & 141 & 173 & 53 & 31 & 35 & 55 & 160 & 135 & 143 & 161 & 566 & 715 & 710 & 599 \\ 
    & 40\% &            177 & 141 & 149 & 178 & 63 & 33 & 39 & 62 & 147 & 120 & 128 & 146 & 604 & 691 & 685 & 589 \\ 
    &&\multicolumn{16}{c}{Scenario (V-C)}\\
    200 & 20\% &  234 & 208 & 207 & 239 &  89 & 71 & 71 & 108 & 157 & 140 & 141 & 174 & 527 & 747 & 733 & 580 \\ 
    & 40\% &            242 & 222 & 223 & 257 & 101 & 80 & 81 & 128 & 145 & 129 & 130 & 167 & 549 & 712 & 694 & 568 \\ 
     400 & 20\% &       219 & 179 & 181 & 226 &  83 & 54 & 55 &  91 & 151 & 125 & 127 & 159 & 581 & 827 & 812 & 609 \\ 
    & 40\% &            219 & 192 & 194 & 237 &  90 & 61 & 62 & 101 & 137 & 114 & 112 & 146 & 631 & 808 & 797 & 595 \\ 
    [1ex]
    \bottomrule
  \end{tabular}
\end{table}

\section{Application to the Cystic Fibrosis Foundation Patient Registry Data}

Understanding the risk factors associated with the progressive loss of lung function is crucial in managing CF. The risk for lung disease depends on patient characteristics, and certain patient groups, such as Hispanic patients, are at increased risk of severe disease for reasons not yet known \citep{mcgarry2019regional}. 
Our goal is to build prediction models for the development of moderate airflow limitation, defined as the first time that ppFEV1 drops below 80\% in CFPPR. Our analysis focused on 5,398 pediatric CF patients who were diagnosed before age one between 2008 and 2013; among them, 419 were Hispanic. The data were subjected to right-censoring due to loss of follow-up or administrative censoring. A total of 4,507 failure events were observed with a median follow-up time of 7.71 years.

The rich information in the CFPPR data renders the possibility of a comprehensive evaluation of important risk factors. We considered baseline predictors including gender, ethnicity, maternal education status ($\ge$16 years of education vs. else), insurance status, geographic location (West, Midwest, Northeast, and South), and mutation class (severe, mild, and unknown). Since baseline factors may have limited predictability, we further included repeated measurements and intermediate events as predictors. The longitudinal measurements, ppFEV1, percent predicted forced vital capacity (ppFVC), weight, and height, were assessed regularly throughout the study and were annualized at integer ages via last observation carried forward. The intermediate events include different subtypes of PA (initial acquisition, mucoid, chronic, multidrug-resistant), methicillin-sensitive staphylococcus aureus (MSSA), methicillin-resistant staphylococcus aureus (MRSA), as well as the diagnoses of CF-related diabetes (CFRD)  and pancreatic insufficiency.

For landmark prediction, we considered the following fixed and random landmark times:
\begin{enumerate}
    \item[(LM1)] The landmark time is age 7, with the target prediction interval $[7,22]$. 
    \item[(LM2)] The landmark time is age 12, with the target prediction interval $[12,22]$.
    \item[(LM3)] The landmark time is the acquisition of chronic \emph{Pseudomonas aeruginosa} (cPA), and the target prediction interval is from the time of acquiring cPA to age 22.
\end{enumerate}
The fixed landmark ages 7 and 12 correspond to middle childhood and preadolescence. The random landmark event cPA was considered because patients with cPA are more likely to develop increased inflammation, leading to an accelerated loss in lung function \citep{kamata2017impact}. The median time to cPA in our dataset was 15.5 years. While PA is a frequent pathogen in cystic fibrosis, early PA is eradicated in the majority of patients through inhaled and intravenous antibiotics \citep{doring2004early,heltshe2018longitudinal}. When PA is not eradicated after initial PA, it converts to chronic PA. Since initial PA is frequently eradicated and does not have the long-term impact on pulmonary function, we chose the landmark time of cPA \citep{harun2016systematic}. A model using initial PA as the landmark is reported in the Supplementary Material.
To perform risk prediction, we used Model (LM1) to obtain the future risk for a patient who is event-free at age 7. An updated prediction can be carried out using Model (LM2) if the patient remains event-free at age 12. Upon converting to cPA, the predicted risk can be updated using Model (LM3). Although landmark prediction models can be constructed at multiple time points, we suggest practitioners focusing on the predicted probabilities given by the landmark model that is most close in time. To illustrate the event history and landmark times, we show the occurrence of PA during the follow-up period for a random sample of 50 patients in the Supplementary Material. At each landmark time, the exact timing of an intermediate event is available only if it has occurred before the landmark time.

The data were partitioned into a training set (60\%) and a test set (40\%). The landmark prediction models were built using the training data and were evaluated on the test data via the proposed concordance measure. Results from the simple landmark Cox model and the two-stage approach were also reported for comparison. Similar to the conventional landmark prediction models, the Cox models were constructed using subjects who remained event-free and uncensored at each landmark time. In the Cox models, the $j$th partially observed intermediate events was incorporated via two predictors, $U_jI(U_j\le\TL)$ and $I(U_j> \TL)$. In the simple landmark Cox model, the partially observed repeated measurements at $t_k$ were expressed using $\bW(t_k)I(t_k \le \TL)$ and $I(t_k > \TL)$ under Model (LM3). 
The models were built in the way described in the simulation section.

The concordance measures are summarized in Table~\ref{tab:con}. For landmark prediction at ages 7 and 12, we reported the average $\CON_t$ at 50 equally spaced time points on the target prediction intervals. 
For the prediction at the landmark age of 12, both Models (LM1) and (LM2) can be applied: (LM1) used history up to age 7 while (LM2) used history up to age 12. As expected, incorporating additional information between ages 7 and 12 results in an increase of average concordance from 0.711 to 0.739 in our ensemble model. For the landmark model at cPA, we used the concordance at a time horizon of 5 years after cPA as the evaluation criterion. Since we focus on the risk prior to age 22, predicting the 5-year risk for individuals who acquired the chronic form of PA after age 17 is not feasible. Therefore, the concordance was evaluated in the subsample of subjects who developed cPA before age 17. The ensemble method yielded better performances compared to its competitors.

\begin{table}[h!]
  \centering
\caption{Concordance measures evaluated using the test data in the CFFPR analysis. The column Landmark gives the left bound of the target prediction interval. The integrated concordance is reported for Models (LM1) and (LM2) over the interval where the risk prediction is performed, where $\widehat\CON_{[a,b]} = \sum_{j=1}^{50}\widehat\CON_{t_j}/50$ and $t_j = a + (b-a)j/50$. The concordance at year 5 after cPA is reported for Model (LM3).}
  \label{tab:con}
  \begin{tabular}{lllcccc}
    \toprule    
    Landmark & Measure & Model & Tree & Ensemble & Cox-1 &  Cox-2 \\ 
    \midrule
    Age 7 & $\widehat\CON_{[0,15]}$ & LM1 & 0.654 & 0.748 & 0.695 & 0.699 \\ 
    \hline
    Age 12 &$\widehat\CON_{[0,10]}$ & LM2 & 0.667 & 0.739 & 0.674 & 0.674 \\ 
    Age 12 & $\widehat\CON_{[0,10]}$ & LM1 & 0.620 & 0.711 & 0.611 & 0.669 \\
    \hline
    cPA & $\widehat\CON_{5}$ & LM3 & 0.763 & 0.813 & 0.788 & -\\
    \bottomrule
  \end{tabular}
\begin{flushleft}
Note: Cox-1 stands for the simple one-stage landmark Cox models, and Cox-2 stands for the two-stage landmark Cox models.
\end{flushleft}
\end{table}

\begin{figure}[h!]
\centering
  \begin{subfigure}[t]{.8\textwidth}
    \centering
    \includegraphics[width = \textwidth]{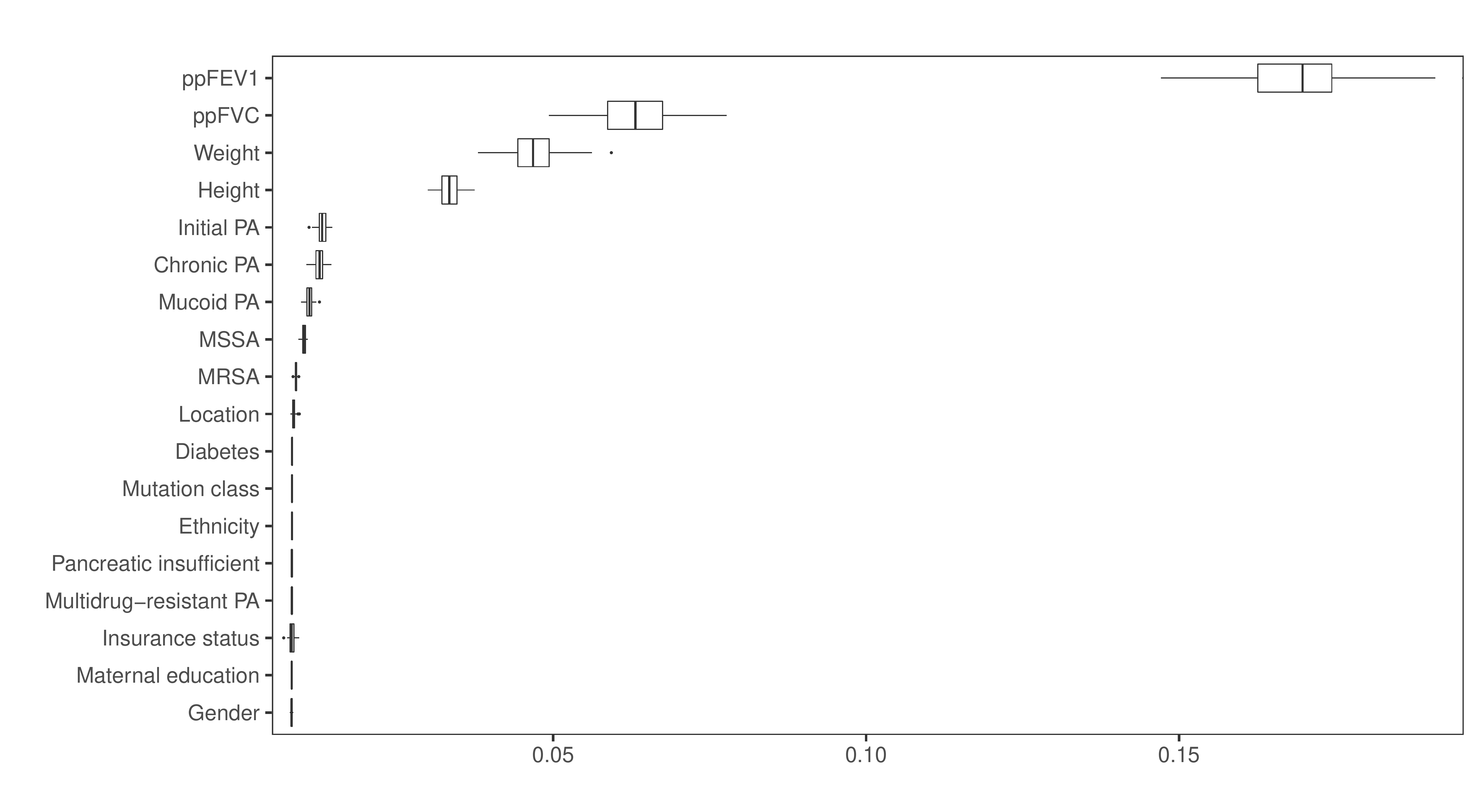}
    \caption{Variable importance when the landmark time is age 7 and the risk prediction interval is $[7, 22]$.}
    \label{fig:vimp3}
  \end{subfigure}\hspace*{.25cm}
  
  \begin{subfigure}[t]{.8\textwidth}
    \centering
    \includegraphics[width = \textwidth]{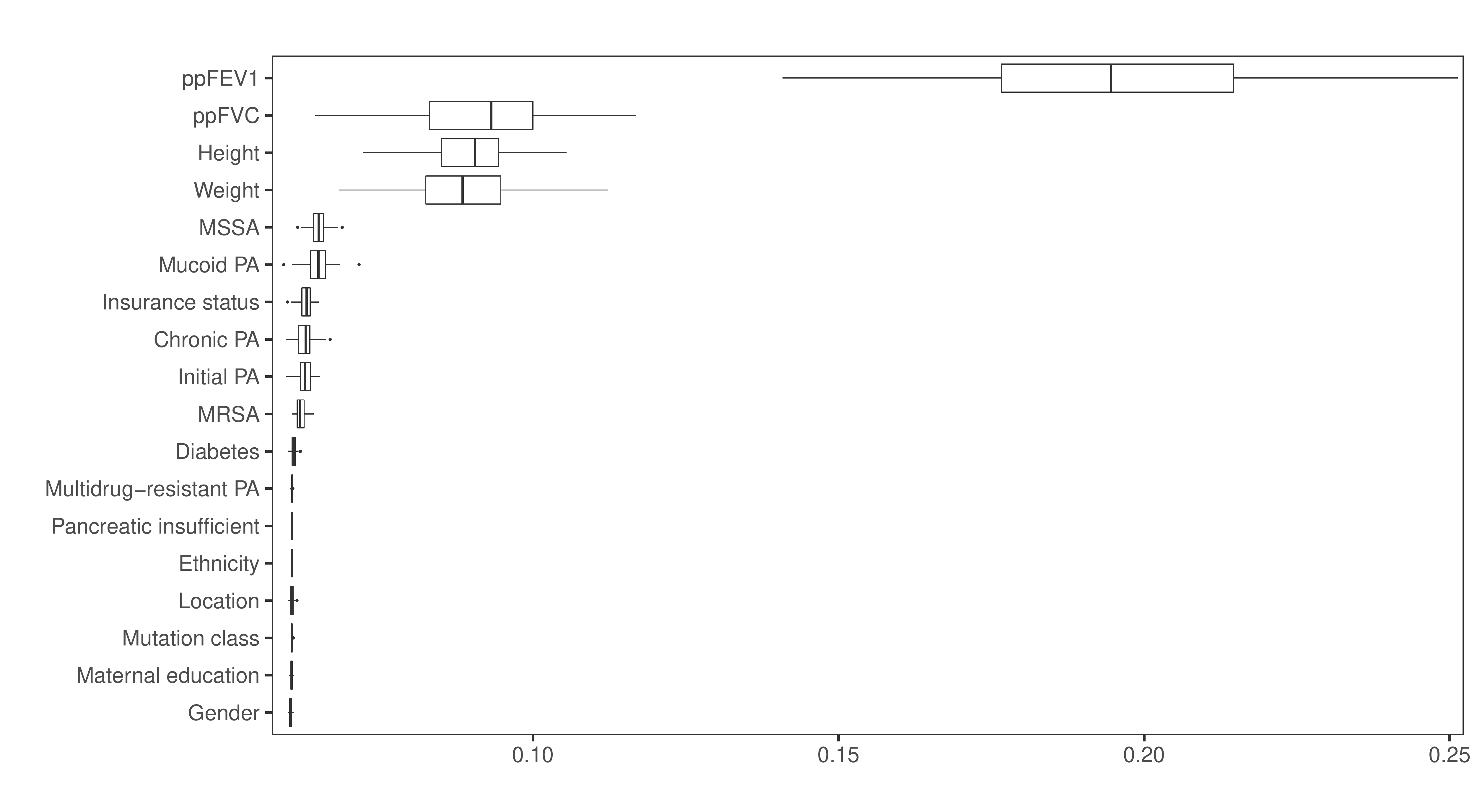}
    \caption{Variable importance when the landmark time is age 12 and the risk prediction interval is $[12, 22]$.}
    \label{fig:vimp2}
  \end{subfigure}

  \caption{The permutation variable importance in the CFFPR data analysis, when the landmark times are ages 7 and 12. The boxplots show the decreases in OOB concordances from the 100 permutations and are ranked in descending order according to the mean value.}
  \label{fig:vimp}
\end{figure}

\begin{figure}[h!]
\centering\vspace*{-1cm}
  \begin{subfigure}[t]{.75\textwidth}
    \centering
    \includegraphics[width = \textwidth]{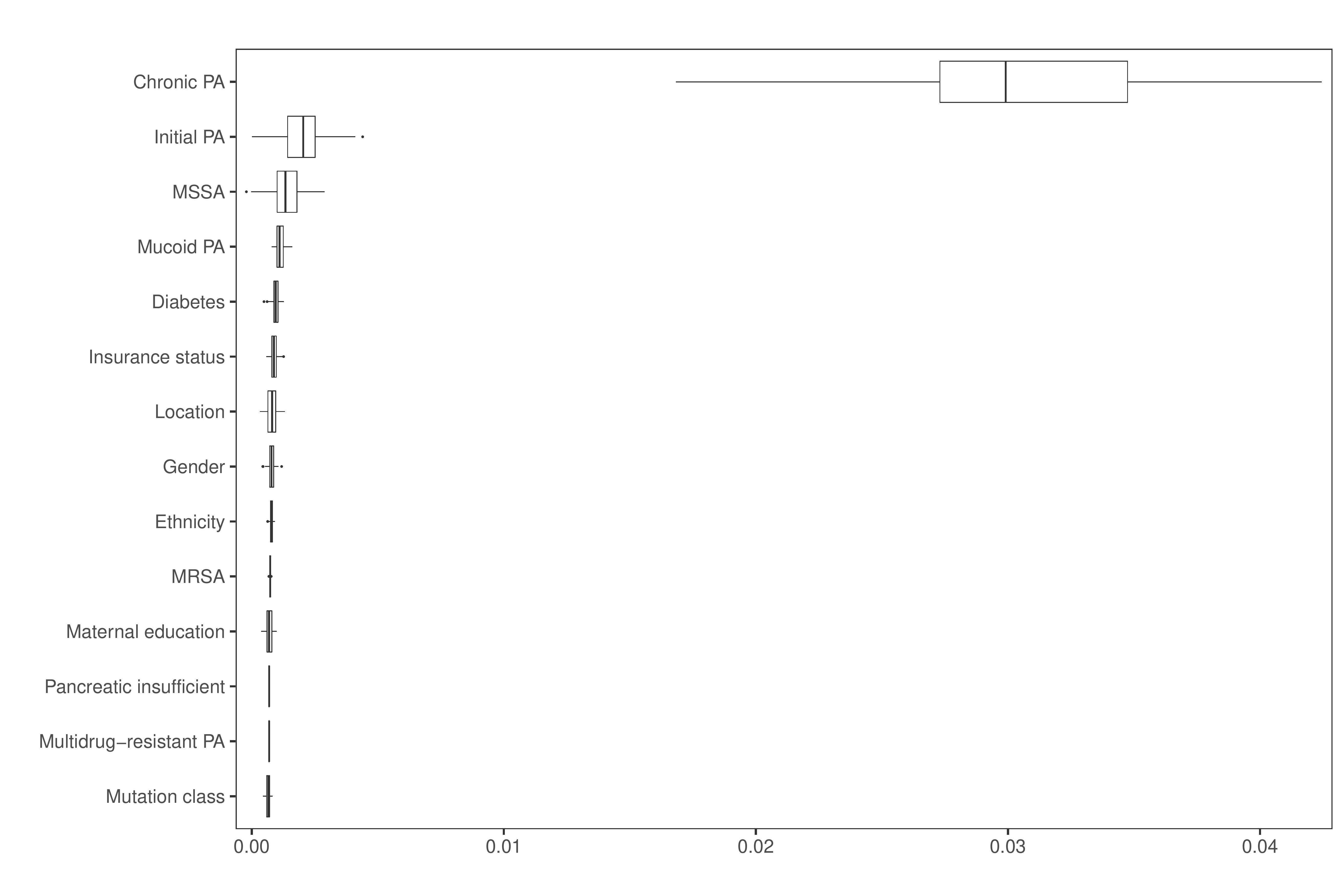}
    \caption{Variable importance for baseline variables and intermediate events}
    \label{fig:vimp-random-b}
  \end{subfigure}\hspace*{.25cm} 
  
  \begin{subfigure}[t]{.45\textwidth}
    \centering
    \includegraphics[width = \textwidth]{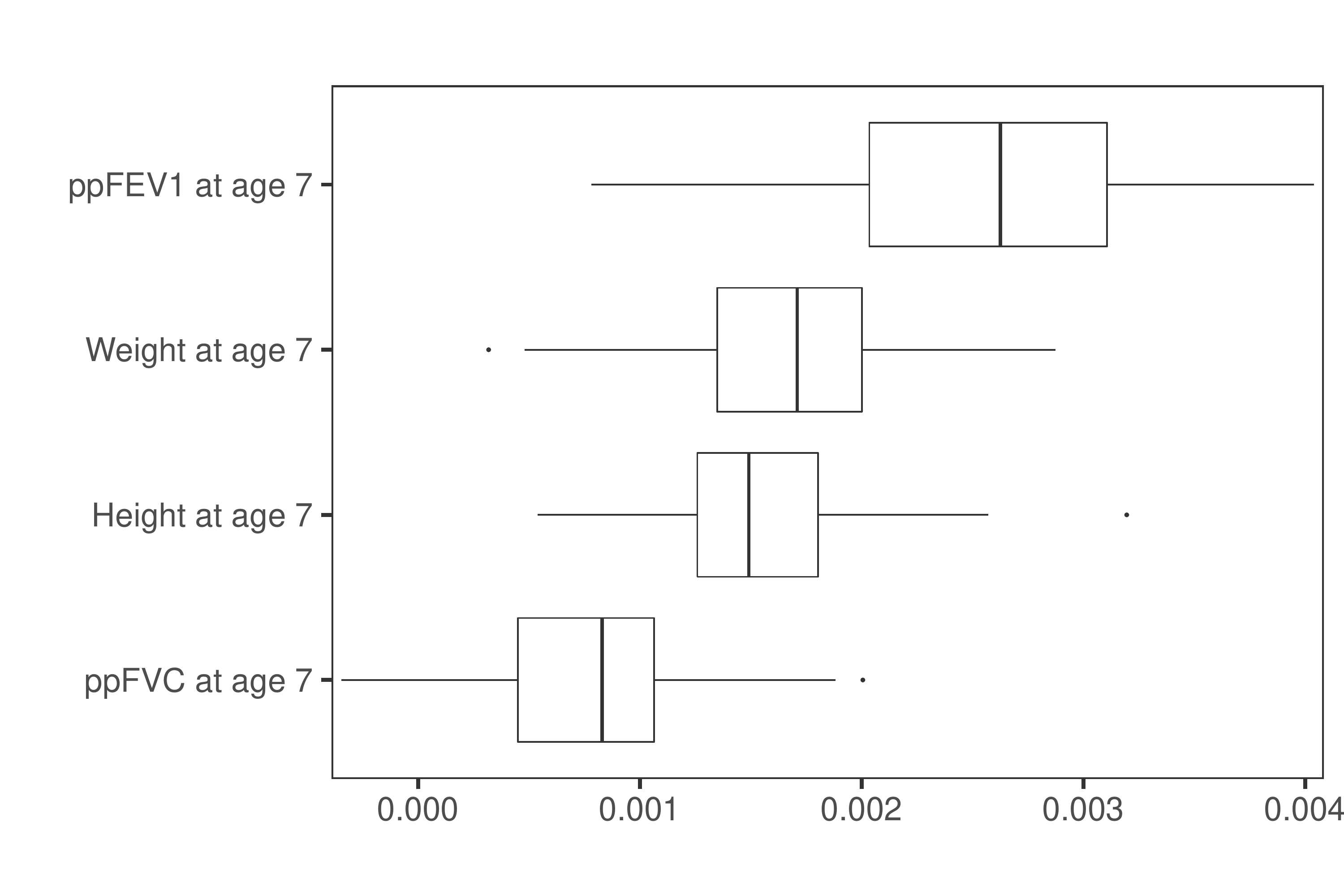}
    \caption{Variable importance for markers  at age 7.}
    \label{fig:vimp-random7}
  \end{subfigure}\hspace*{.25cm}
  \begin{subfigure}[t]{.45\textwidth}
    \centering
    \includegraphics[width = \textwidth]{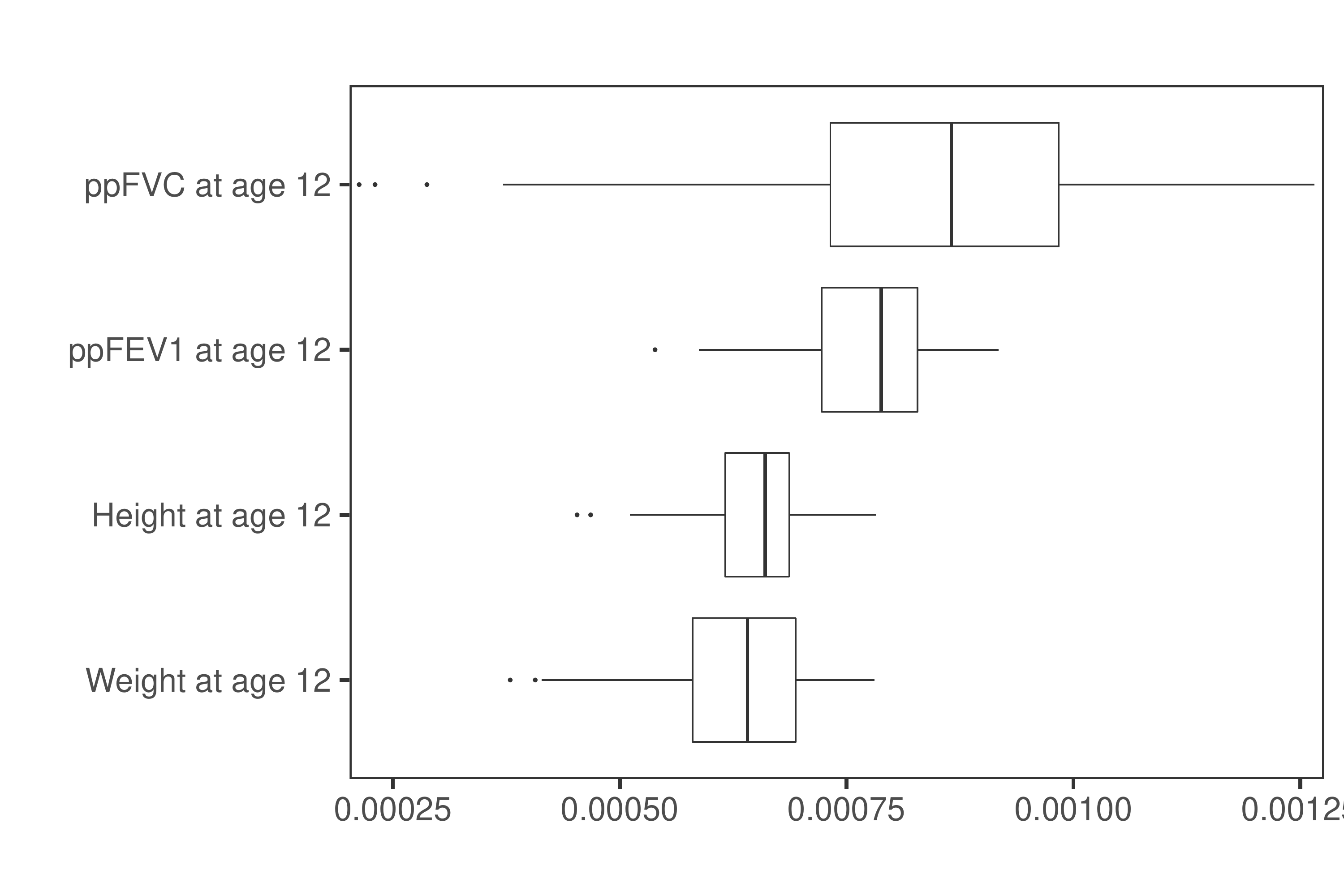}
    \caption{Variable importance for markers at age 12.}
    \label{fig:vimp-random12}
  \end{subfigure}\hspace*{.25cm}  
   \caption{The permutation variable importance in the CFFPR data analysis, when the landmark time is cPA. The boxplots show the decreases in OOB concordances from the 100 permutations and are ranked in descending order according to the mean value.}
\label{fig:vimp-random}
\end{figure}

In an attempt to identify important predictors in the ensemble models, we computed the permutation variable importance with 100 permutations. Under the fixed landmark time models, we permuted all of the repeated measurements of a marker simultaneously to evaluate the overall impact of the longitudinal marker. The results of the permutation variable importance are summarized in Figure~\ref{fig:vimp}, where ppFEV1, ppFVC, weight, and height are identified as the top four important predictors for both landmark ages 7 and 12 (Figures~\ref{fig:vimp3} and \ref{fig:vimp2}). Following them, intermediate events related to PA and staphylococcus aureus are moderately important. We note that mucoid PA and MSSA became more important at age 12 when compared to age 7. This could be due to the fact that these intermediate events are less common before age 7. When using cPA as the landmark, the repeated measurements after the acquisition of cPA were not used in prediction, and thus the number of observed repeated measurements varied across subjects. Unlike baseline variables and intermediate events of which the permutations were performed among subjects who experienced cPA, the permutation of a marker at a specific time point (e.g., age 7) was performed among subjects who experienced cPA after the time point. To this end, we plot the variable importance for predictors at different ages separately, so that the variable importance measures in each plot are based on permuting the same set of subjects. Figure \ref{fig:vimp-random-b} shows the importance of baseline variables and intermediate events. The timing of cPA plays an important role in predicting future event risk. 
We note that baseline factors such as ethnicity have a relatively low variable importance. However, this does not mean that ethnicity does not affect the risk of lung function decline. We conjecture that the effect of ethnicity was predominantly mediated through spirometry measurements. Therefore, one should be cautious when interpreting the variable importance. When applying the proposed method in health disparity research, one can further build separate models for Hispanic and non-Hispanic patients. Additional analyses in different ethnic groups are included in the Supplementary Material.

\begin{figure}[h!]
  \centering
  \begin{subfigure}[t]{\textwidth}
    \includegraphics[width = \textwidth]{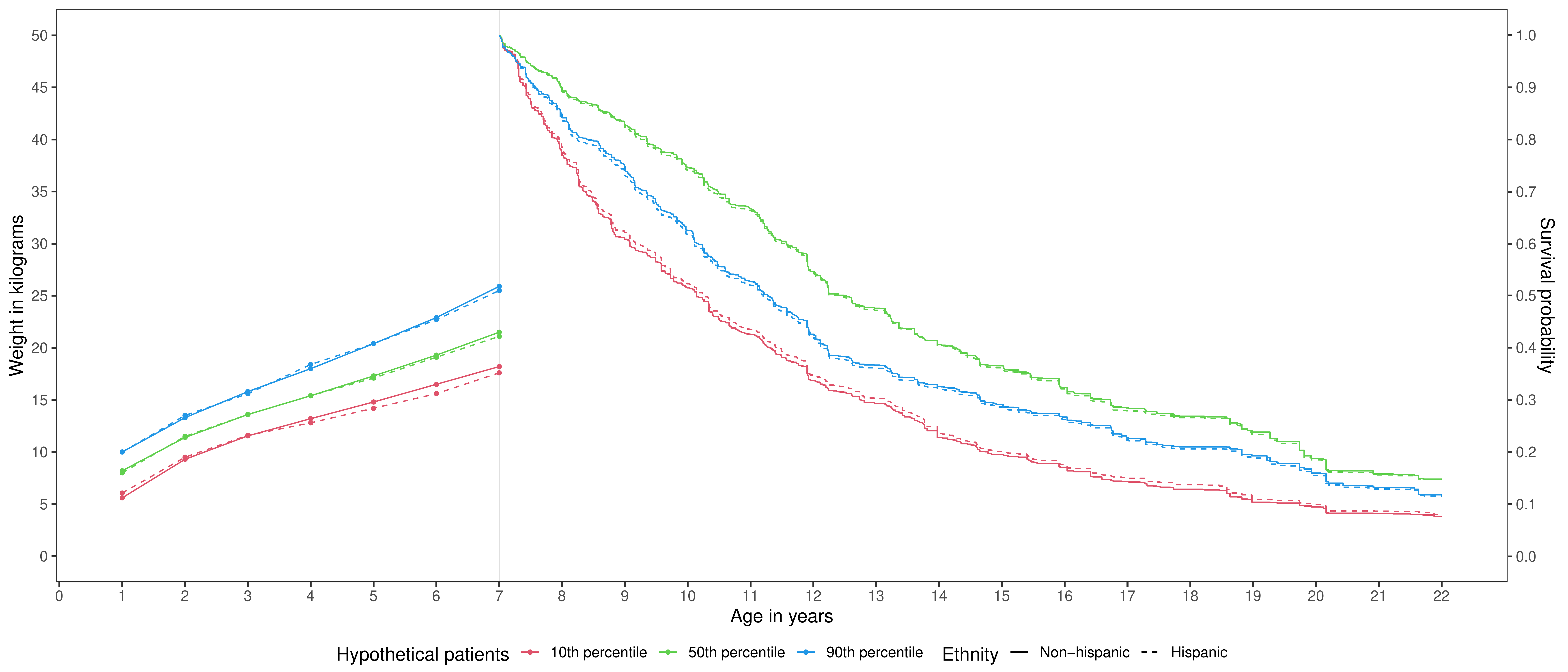}
    \caption{Landmark time point at age 7, predicting the event risk on $[7, 22]$.}
    \label{fig:surv2}
  \end{subfigure}
  \begin{subfigure}[t]{\textwidth}
    \includegraphics[width = \textwidth]{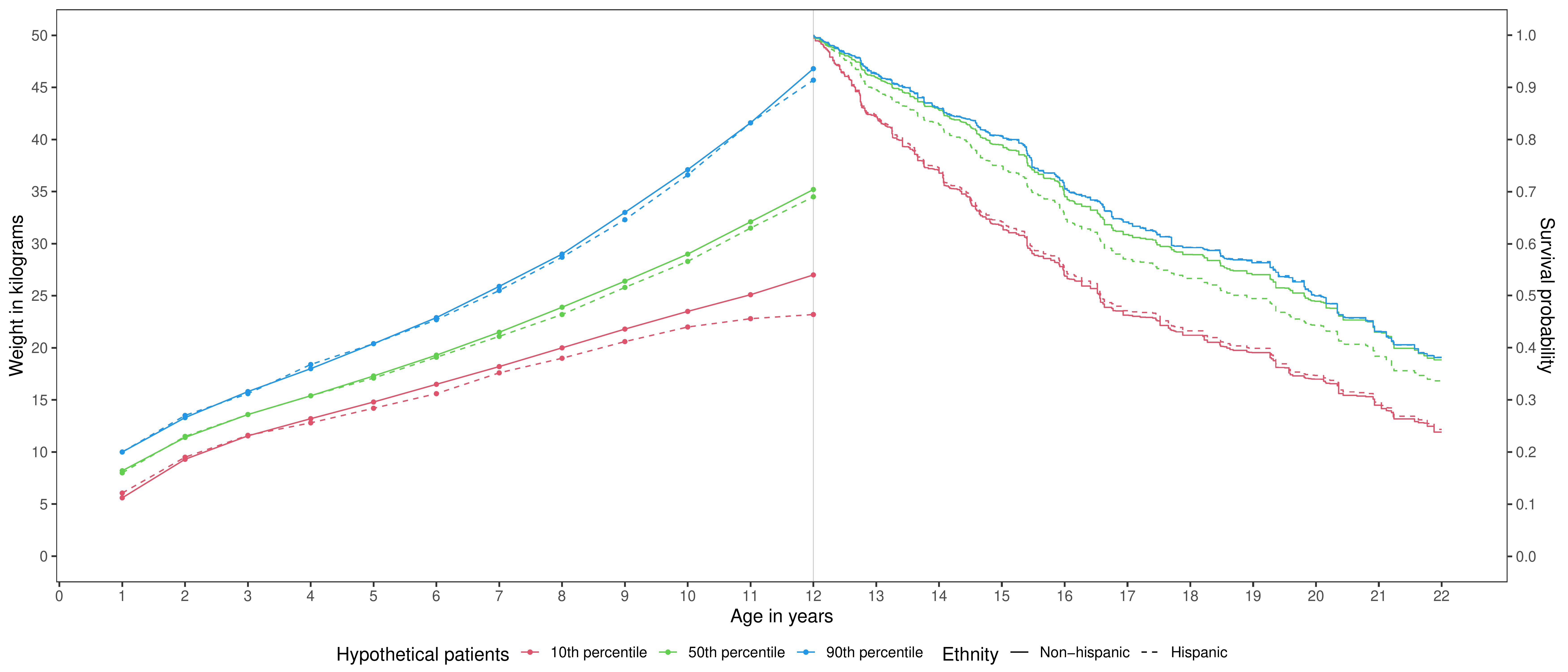}
    \caption{Landmark time point at age 12, predicting the event risk on $[12, 22]$.}
    \label{fig:surv3}
  \end{subfigure}
  \caption{Survival predictions for patients in different weight groups.
    {Curves on the left show} the repeated weight measurements before the landmark time.
    {Curves on the right show} the survival predictions over the interval of interests for the
    corresponding groups. The weight groups were chosen by the percentiles and ethnicity,
    \lineCir[red!90]{.1cm}: 10th percentile, non-Hispanic; \lineCir[green!90]{.1cm}: 50th percentile, non-Hispanic; \lineCir[blue!90]{.1cm}: 90th percentile, non-Hispanic; 
    \dashlineCir[red!90]{.1cm}: 10th percentile, Hispanic; \dashlineCir[green!90]{.1cm}: 50th percentile, Hispanic; \dashlineCir[blue!90]{.1cm}: 90th percentile, Hispanic. }
    \label{fig:surv}
\end{figure}

Our models identify repeated measurements of weight and height as important variables in landmark prediction.
To provide more insight into how historical weight measurements affect future risk in our ensemble model, we present the predicted event-free probabilities for hypothetical patients with different weight trajectories in Figure~\ref{fig:surv}. Specifically, we consider Hispanic and non-Hispanic male patients whose weights were in the 10th, 50th, and 90th weight-for-age percentiles of the corresponding ethnicity-gender subgroup. For all six patients, the intermediate event times were fixed at the median derived from the Kaplan-Meier estimates, while categorical predictors and continuous predictors were fixed at the reference levels and mean values, respectively. At the landmark age 7, patients with the 10th percentile weight trajectories had the highest predicted risk, followed by patients with 90th percentile and those with 50th percentile weight trajectories (Figure~\ref{fig:surv2}). 
At the landmark age 12, the predicted risk in patients with 10th percentile weight remains the highest, followed by patients with 50th percentile and those with 90th percentile weight trajectories (Figure~\ref{fig:surv3}). A possible explanation for the predicted curves of the 50th and 90th weight percentiles to flip between landmark age 7 and 12 is a higher degree of survivor bias at the later landmark age; in other words, the event-free individuals at landmark age 12 tended to have better lung function than those at landmark age 7. So if overweight is associated with a higher risk of lung function decline, the 90th percentile patients are more likely to fail before age 12. As a result, we have a group of healthier overweight subjects at the landmark age 12, and thus their risk beyond age 12 can be lower than individuals with 50th percentile weight. As for underweight individuals, the degree of selection bias may not be large enough to compensate for the poor lung function and thus remains the highest risk group at landmark age 12. To summarize, the landmark prediction models built at age 7 and age 12 are for different survivor populations, and thus can not be directly compared.

\section{Discussion}
In this paper, we proposed a unified framework for tree-based risk prediction with updated information. Compared to semiparametric methods, our methods can handle a large, growing number of predictors over time and do not impose strong model assumptions. Furthermore, the landmark times at which a prediction is performed are allowed to be subject-specific and defined by intermediate clinical events.  Notably, our ensemble procedure averages the unbiased martingale estimation equations instead of survival probabilities and avoids the potential bias arising due to small terminal node sizes.

Our discussion has focused on the case where the time-dependent variables $\bW(\cdot)$ are observed at fixed time points $t_1,\ldots,t_K$. The proposed method can also be applied to the case where repeated measurements are collected at irregular time points, such as hospitalizations. 
When building prediction models, one can consider using up to $K$ repeated measurements as predictors, where $K$ is a fixed integer. Denote by $V_1<V_2<\cdots<V_K$ the potential observation times of $\bW(\cdot)$.
At time $t$, the available information can be expressed using $\bX(t) = \{\bW(V_1,t), \ldots, \bW(V_K,t), V_1(t), \ldots, V_K(t)\}$, where $\bW(V_k,t) = \bW(V_k)$ and $V_k(t) = V_k$ if $V_k \le t$, while $\bW(V_k,t) = \textbf{NA}_q$ and $V_k(t) = t^+$ otherwise. In other words, the random measurement times $V_1,\cdots,V_K$ are treated as intermediate event times. 
In this way, our framework can incorporate irregularly observed covariate information.

\section*{Acknowledgements}

The authors would like to thank the three anonymous referees, the Associate Editor, and the Editor for their helpful comments that improved the quality of this paper. The authors thank the CF Foundation for the use of the CFFPR
data to conduct this study. Additionally, we would like to thank the patients, care providers, and care coordinators at CF centers throughout the U.S. for their contributions to the CFFPR. The views expressed in this manuscript are those of the authors and do not necessarily represent the views of the National Heart, Lung and Blood Institute (NHLBI), the National Institutes of Health (NIH) and the U.S. Department of Health and Human Services. Sun's research was supported by NIH (R21HL156228). Wu's research was supported in part by the Intramural Research Program of the NHLBI/NIH. Huang's research was supported by NIH (R01CA193888). McGarry's research was supported by NIH (1K23HL133437-01A1) and CF Foundation Therapeutics (MCGARR16A0).

\bibliographystyle{asa} 
\bibliography{ref}

\end{document}